\documentclass{WileyMSP-template}

\usepackage{textcomp}
\usepackage{gensymb}
\usepackage{amsmath}
\usepackage{booktabs}
\usepackage{multirow}
\usepackage{microtype}
\usepackage{makecell}
\usepackage[english]{babel}  % or your language
\usepackage[table]{xcolor}  % needed for coloring rows
\usepackage{csquotes}   % Handles quotes properly
\usepackage[
  backend=biber,
  style=numeric,
  sorting=none,
  doi=true,
  url=false,
  isbn=true
]{biblatex}

\usepackage{hyperref}

\hypersetup{
  colorlinks=true,
  linkcolor=black,   % internal links (sections, refs)
  urlcolor=blue,   % URLs
  citecolor=black  % citations
}

\definecolor{lobcolor}{HTML}{a6cee3}  % replace 1F77B4 with your hex

\begin{document}

\pagestyle{fancy}
\fancyhf{}
\fancyfoot[C]{\thepage}
%\rhead{\includegraphics[width=2.5cm]{vch-logo.png}}

\title{A critical assessment of bonding descriptors for predicting materials properties}

\maketitle

% Author: Please give full first and last names for authors and include * after the name of all corresponding authors

\author{Aakash Ashok Naik}
\author{Nidal Dhamrait}
\author{Katharina Ueltzen}
\author{Christina Ertural}
\author{Philipp Benner}
\author{Gian-Marco Rignanese}
\author{Janine George*}

% Dedication

%\dedication{Optional dedication here. If no dedication is required, please leave blank}

% Affiliations: Please provide academic titles (Prof. or Dr.) for all authors where applicable, and include an institutional email address for all corresponding authors
\begin{affiliations}
\noindent Aakash Ashok Naik, Katharina Ueltzen, Prof. Dr. Janine George \\
Address \\
Department of Materials Chemistry, Federal Institute for Materials Research and Testing, Berlin, 12205, Germany \\
Institute of Condensed Matter Theory and Optics, Friedrich Schiller University Jena, Jena, 10587, Germany \\
Email Address: janine.george@bam.de \\
\hfill \break
Nidal Dhamrait, Dr. Christina Ertural \\ % affiliation of the time when the contributions were made
Address \\
Department of Materials Chemistry, Federal Institute for Materials Research and Testing, Berlin, 12205, Germany \\
\hfill \break
Dr. Philipp Benner \\
Address \\
eScience Group, Federal Institute for Materials Research and Testing, Berlin, 12205, Germany \\
\hfill \break
Prof. Dr. Gian-Marco Rignanese \\
Address \\
Institute of Condensed Matter and Nanosciences (IMCN), UCLouvain, Louvain-la-Neuve, 1348, Belgium \\

\end{affiliations}

% Keywords: Please provide a minimum of three and a maximum of seven keywords, separated by commas


% Abstract should be written in the present tense and impersonal style (i.e., avoid we), and be at most 200 words long
\begin{abstract}

Most machine learning models for materials science rely on descriptors based on materials compositions and structures, even though the chemical bond has been proven to be a valuable concept for predicting materials properties. Over the years, various theoretical frameworks have been developed to characterize bonding in solid-state materials. However, integrating bonding information from these frameworks into machine learning pipelines at scale has been limited by the lack of a systematically generated and validated database. Recent advances in high-throughput bonding analysis workflows have addressed this issue, and our previously computed Quantum-Chemical Bonding Database for Solid-State Materials was extended to include approximately 13,000 materials. This database is then used to derive a new set of quantum-chemical bonding descriptors. A systematic assessment is performed using statistical significance tests to evaluate how the inclusion of these descriptors influences the performance of machine-learning models that otherwise rely solely on structure- and composition-derived features. Models are built to predict elastic, vibrational, and thermodynamic properties typically associated with chemical bonding in materials. The results demonstrate that incorporating quantum-chemical bonding descriptors not only improves predictive performance but also helps identify intuitive expressions for properties such as the projected force constant and lattice thermal conductivity via symbolic regression.

\end{abstract}

% Text: Please use section headings and subheadings as specified below. For communications, all section headings apart from Experimental Section should be removed
% Please make the first reference to a display item bold: \textbf{Figure 1}
% Do not abbreviate Figure, Equation, etc.; display items are always singular, i.e., Figure 1 and 2.
% Equations are always singular, i.e., Equation 1 and 2, and should be inserted using the {equation} environment, not as graphics
% Please do not use footnotes in the text, additional information can be added to the Reference list.

\section{Introduction}

Machine learning algorithms are now widely used in data-driven materials discovery, both in forward and inverse design approaches.\cite{himanen2019data, himanen2019datacorrection, noh2020machine} In the context of forward design, where the main goal is to predict material properties based on structure or composition as input, the performance of the employed machine learning algorithms depends to a large extent on how well the materials are represented. This representation is generally referred to as a set of features or descriptors. These descriptors are typically derived from a material's crystal structure, composition, or electronic properties.\cite{seko2017, ramprasad2017machine, ward2018matminer, jenke2018electronic, tawfik2022naturally} Numerous studies have already demonstrated the utility of these descriptors for building machine learning models that can be used for screening materials of interest for various applications, such as catalysis,\cite{toyao2019machine} ferroelectrics,\cite{he2021machine}, and thermoelectrics,\cite{carrete2014finding}, among others.\cite{NEMATOV2025e01139}

The concept of a chemical bond, while not a quantum mechanically observable quantity, has proven helpful in rationalizing both organic and inorganic materials, whether it is the reaction pathways in organic compounds,\cite{woodward1965stereochemistry} the nature of magnetic superexchange interactions between transition metal ions,\cite{goodenough1963magnetism}, or crystal structure stability.\cite{goldschmidt1926gesetze, pauling1929principles} To characterize bonding within solid-state materials, several theoretical frameworks have been developed. These include wavefunction- and population-analysis,\cite{mulliken1955electronic, dunnington2012generalization, galeev2013solid} real-space electron density analysis,\cite{bader1990quantum, savin1992electron, bagus1997charge} and energy partitioning methods.\cite{raupach2015periodic, kreuter2025energy}
These frameworks are routinely employed to understand or tune various material properties.\cite{schon2022classification, ray2022programming, zhou2024chemical, ubaid2024antibonding, powell2025exploiting}  The quantities obtained through such bonding analysis frameworks can therefore also serve as valuable descriptors for data-driven material discovery.

To date, descriptors derived from readily available geometric information have often been used to approximately characterize bonding in materials.\cite{seko2017, botu2015adaptive, waroquiers2017statistical, waroquiers2020chemenv, musil2021physics} Currently, a large-scale comparison of the predictive power of geometric descriptors and quantum-chemical bonding descriptors in machine learning of material properties for solid-state materials is lacking. Recent developments in quantum-chemical bonding analysis workflows have enabled the high-throughput computation of quantum-chemical bonding descriptors derived from ab initio calculations.\cite{george2022automated, ganose2025atomate2}
Based on these developments, a database of chemical bonding descriptors for 1500 semiconductors \cite{naik2023quantum, naik_2023_7852083} was published by some of us. As part of the current study, we have further extended this database, which now includes bonding properties for approximately 13,000 materials from the Materials Project.\cite{jain2013materials} The materials were filtered according to the availability of computed elasticity data in the Materials Project database version 2023.11.1. An overview of the structure and composition coverage in the bonding analysis dataset is presented in Supplementary Information Section \ref{database_overview} (Figure \ref{fig:overview_1} and \ref{fig:overview_2}), demonstrating that our curated dataset is highly diverse and not limited to any specific class of materials in terms of composition, structure, or bonding. This database consists of bonding indicators generated using the LOBSTER \cite{deringer2011crystal, maintz2013analytic, maintz2016lobster,nelson2020lobster} program, a projection-based bonding analysis framework. Specifically, it includes Crystal Orbital Overlap Populations (COOP),\cite{hughbanks1983chains}  Crystal Orbital Hamilton Populations (COHP),\cite{dronskowski1993crystal}, and two-center Crystal Orbital Bond Index (COBI),\cite{mueller2021crystal}, along with their energy-integrated counterparts up to the Fermi level: ICOOP, which measures the number of electrons participating in a bond; ICOHP, which quantifies covalent bond strength; and ICOBI, which indicates bond order. In addition, Mulliken and Löwdin atomic charges,\cite{ertural2019development} projected densities of states (PDOS), and Madelung energies \cite{trefry1987electrostatic,hoppe1995madelung} are also available.

With this large database at hand, we evaluate the predictive value of descriptors derived from such bonding indicators for data-driven materials discovery. Since these bonding indicators have not been comprehensively assessed before in the context of data-driven materials science, this study focuses on statistical descriptors derived from COHPs, ICOHPs, and atomic charges (Mulliken/L{\"o}wdin). These descriptors are extracted using LobsterPy \cite{naik2024lobsterpy}, a Python package that provides tools for automatically generating summaries of bonding characteristics in materials and transforming data from LOBSTER outputs into machine-learning-ready formats. Because these descriptors quantify interatomic interactions, they are closely linked to vibrational properties, which are governed by the interatomic force constants. Accordingly, the target material properties considered in this study include the maximum bond-projected force constant, the last peak (highest-frequency) of the phonon density of states (DOS), thermodynamic data (such as heat capacity, vibrational entropy, Helmholtz free energy, and internal energy), mean squared thermal displacements, elasticity data (bulk and shear modulus), and lattice thermal conductivity.  
The rationale for selecting these specific targets is as follows: The bond-projected force constant, which measures bond stiffness, has been shown to be useful for detecting long-range interactions \cite{hempelmann2021long} and for screening two-dimensional materials.\cite{bagheri2023identification} The last peak of the phonon DOS, which is a benchmark property in Matbench for evaluation of machine learning models,\cite{dunn2020benchmarking}, by definition, is close to the highest vibrational frequency within a material. This frequency is generally indicative of the strongest bond present in the material \cite{gong2024machine} and can also be used to obtain insights into the critical temperature $T_c$ in superconducting materials.\cite{trachenko2025upper} The correlation between the strongest bond and the last peak of the phonon DOS has been demonstrated in a pilot study involving bonding descriptors conducted by some of us.\cite{naik2023quantum} The thermodynamic properties, bulk/ shear modulus, mean squared displacements, and lattice thermal conductivity in materials are also commonly correlated or explained through concepts of chemical bonding.\cite{zeier2016thinking, he2022accelerated, isotta2023elastic}. Additionally, the bonding descriptors evaluated in this study are orders of magnitude less expensive to compute than target properties such as phonons, elastic moduli, or thermal conductivities using standard density functional theory (DFT) simulations.

Through this evaluation, we seek to address three primary questions: (a) Are such quantum-chemical descriptors relevant for predicting these material properties? (b) Can such bonding descriptors be replaced by descriptors derived from compositional and structural data in specific or all cases? (c) Do quantum-chemical chemical bonding descriptors contain complementary information that enhances predictive accuracy beyond simple compositional or structural descriptors? Therefore, we begin by testing the relevance of our quantum-chemical descriptors for learning these properties, i.e., their ability to provide meaningful information for learning the target property of interest. Next, we analyze the correlations between relevant quantum-chemical bonding descriptors and structural or compositional descriptors to gain initial insight into whether our descriptors offer any complementary information that could improve the learning of target properties. We then assess the impact of including these descriptors on the predictive performance of machine learning models, specifically in the Random Forest and MODNet \cite{de2021materials, de2022accurate} models. Significance tests are conducted on trained models to determine whether the observed improvement, if any, is statistically significant. Descriptor importance from the trained models is extracted using explainable artificial intelligence (XAI) techniques, specifically Shapley additive explanations (SHAP) \cite{lundberg2017unified, lundberg2020local} and permutation feature importance (PFI),\cite{altmann2010permutation} to identify which descriptors are most influential. When including bonding descriptors improves predictive performance for a material property, we apply the symbolic regression method SISSO \cite{ouyang2018sisso, Purcell2022, purcell2023recent} to explore whether simple, intuitive expressions composed of these descriptors can be found that relate to that property. In doing so, we aim to gain deeper insight into the connection between bonding descriptors and material properties.

\begin{figure}[!h]
\includegraphics[width=0.95\linewidth]{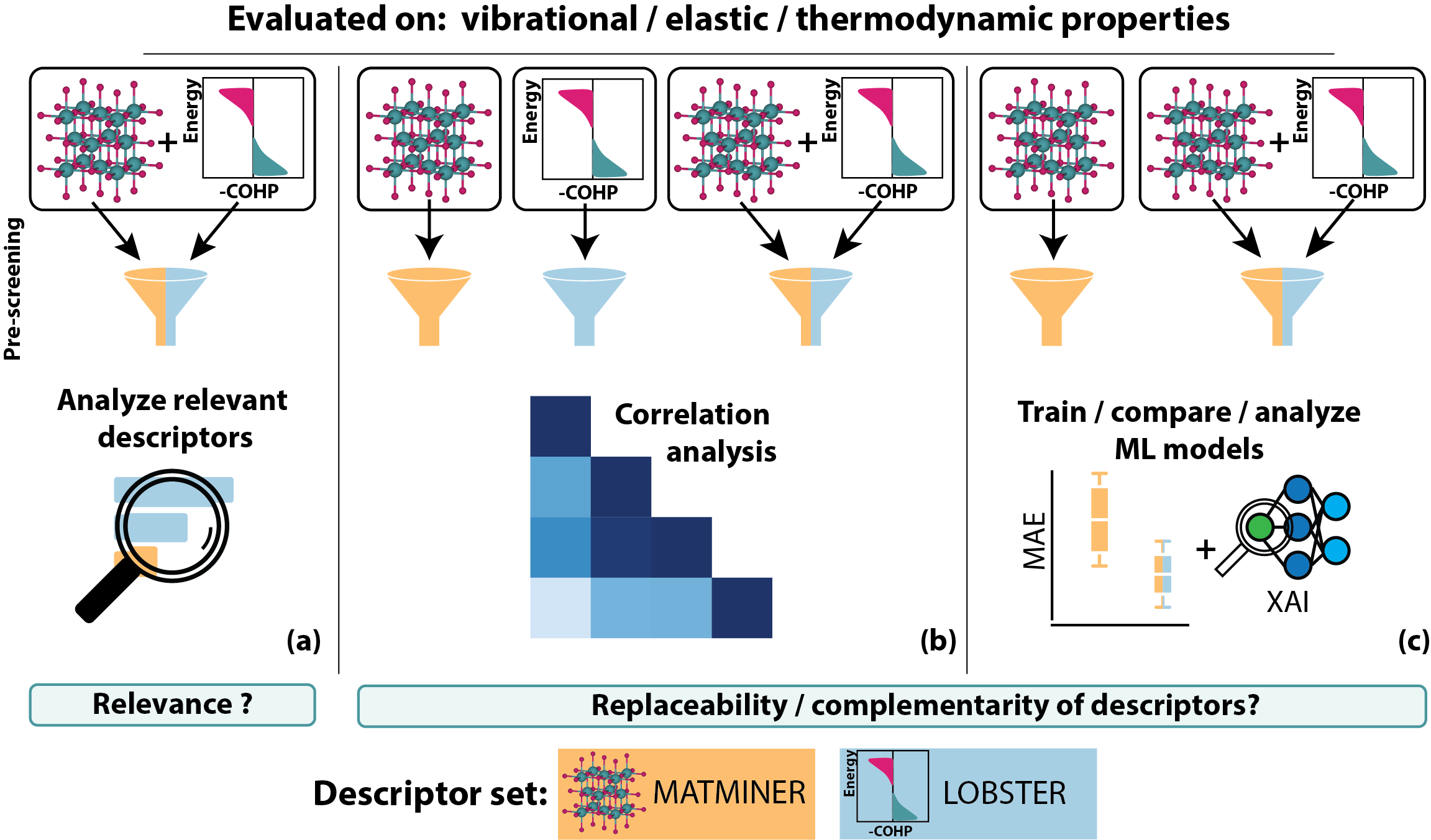}
\caption{Simplified overview of the methods employed for the bonding descriptor evaluation. The color (symbol) of the bottom legend distinguishes descriptor groups derived from ``MATMINER" (structure- and composition-based) and ``LOBSTER" (bonding descriptors from LOBSTER calculations). The funnel icons represent the step related to pre-screening the relevant descriptors; details of this step are shown in Figure \ref{fig:mlpipeline}. (a) The most relevant descriptors identified are analyzed to address the question of their relevance. (b) Correlation analysis to evaluate redundancy and interdependence among descriptor sets. (c) Evaluation of the impact of descriptors on machine learning models' performance, combined with explainable AI (XAI) methods. Together, (b) and (c) address the question of replaceability and complementarity of the evaluated descriptor sets.}\label{fig:overview_schematic}
\end{figure}

\section{Results and Discussions}

We evaluated our descriptors using multiple methods and across a variety of target material properties. Figure \ref{fig:overview_schematic} shows a simplified overview of the employed descriptor evaluation methods. For clarity and brevity, we focus our discussion on a small number of representative methods and targets from each sub-method depicted in Figure \ref{fig:overview_schematic} that best capture the study's main conclusions. These examples include both cases where the descriptors perform well and cases where they offer little to no improvement. Full methodological details are provided in Section~\ref{methods}, and the complete set of results for all methods and targets is available on the repository's \href{https://digimatchem.github.io/paper-ml-with-lobster-descriptors/}{GitHub pages}, organized for easy access and navigation.

\subsection{Relevant descriptors}\label{rel_des_results}

\begin{figure}[ht]
\includegraphics[width=0.90\linewidth]{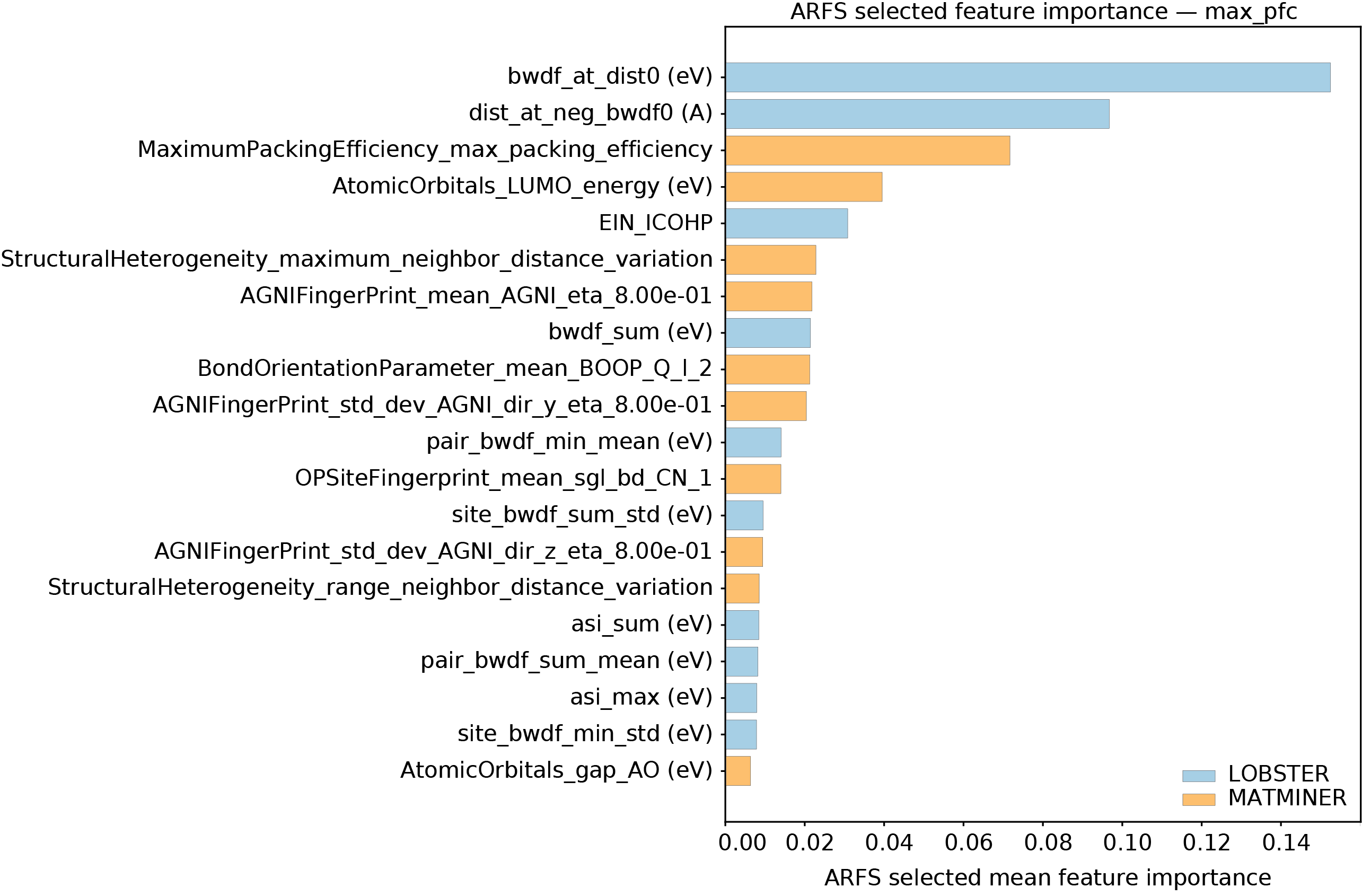}
\caption{Descriptor ranking based on ARFS scores for the maximum of bond-projected force constant (max\_pfc), where the legend distinguishes descriptor groups derived from ``MATMINER" (structure- and composition-based) and ``LOBSTER" (extracted from LOBSTER calculation data).}\label{fig:arfs_feat_imp}
\end{figure}

To avoid overfitting, we perform an initial descriptor selection. The importance scores shown in Figure \ref{fig:arfs_feat_imp} correspond to this pre-processing step in our descriptor evaluation pipeline. We only show the average feature importance scores for ``max\_pfc" as an illustrative example. The descriptors are ranked by importance scores from the all‑relevant feature selection (ARFS) method. Further details are summarized in Section \ref{rel_desc} with additional information provided in the Supplementary Information Section \ref{rel_desc_details}. For the examined targets, we find that primarily the following bonding descriptors are of relevance (i.e., they are selected as part of the feature selection procedure): bond strengths (*\_bwdf\_sum, Icohp\_sum), effective coordination numbers (EIN\_ICOHP), the local environment asymmetry index (asi\_*), atomic charge-based statistics (Loewdin\_*, Mulliken\_*), and bonding heterogeneity (*\_bwdf\_skew*, *\_bwdf\_kurtosis*). Further detailed information on the descriptor formulations and notations can be found in Supplementary Information Section \ref{desc_formulations} and Table \ref{tab:bonding_descriptors}, respectively. A high ranking of these descriptors, appearing in the top 20 according to the feature selector, is observed only for the cases of maximum of bond-projected force constants (max\_pfc), last phonon DOS peak (last\_ph\_peak), average Total/Peierls lattice thermal conductivity (log\_klat\_300/ log\_kp\_300), bulk/shear modulus (log\_k\_vrh/ log\_g\_vrh), and mean-squared displacement (log\_msd\_*). 

Our statistical bonding descriptors rank relatively low for the vibrational part of the thermodynamic properties. These include Helmholtz energy (H\_25, H\_305, H\_705), vibrational entropy (S\_25, S\_305, S\_705), internal energy (U\_25, U\_305, U\_705), and heat capacity (Cv\_25, Cv\_305, Cv\_705), with the subscripts denoting temperatures of 25 K, 305 K, and 705 K.
Instead, we find that average bond length, geometry-based local environment descriptors, and element-based properties, such as atomic weight and covalent radius, are more informative. This suggests that incorporating quantum-chemical bonding descriptors, at least in their current form, into machine learning models for these thermodynamic properties is unlikely to improve predictive accuracy. For other target properties, where our descriptors are ranked higher, we expect to see improvements. To further validate these hypotheses based purely on feature selectors' importance scores, we conduct a correlation analysis on the descriptor sets before training the machine learning models.

\begin{figure}[htbp]
    \centering
    \includegraphics[width=0.7\linewidth]{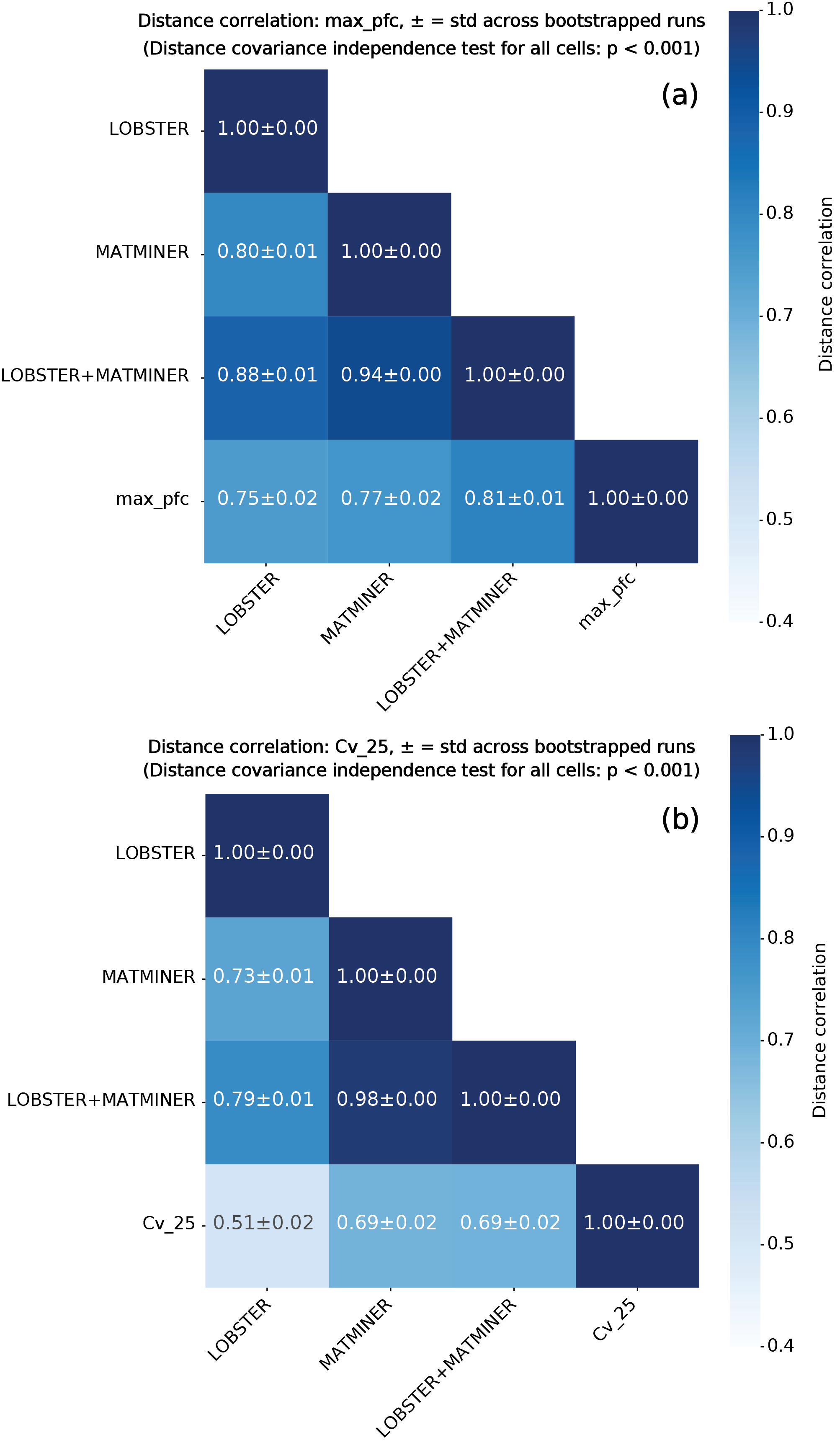}
    \caption{Distance correlation between descriptor sets and targets for (a) Maximum of bond-projected force constant, max\_pfc and (b) Heat capacity at 25 K, Cv\_25. Descriptor sets are ``MATMINER" (structure and composition), ``LOBSTER" (extracted from LOBSTER calculation data), and their combination ``MATMINER+LOBSTER". Only the descriptors relevant to each target are included from each set in this analysis.}
    \label{fig:distance_corr}
\end{figure}

\subsection{Descriptor set and Target correlation analysis}\label{corr_analysis_results}

To shed further light on the relationship between descriptors and targets, we perform a correlation analysis between the descriptor sets and the target property using two methods: one is based on distance correlation, and the other on dependency graphs. 
The distance correlation analysis method is model-agnostic and measures both linear and nonlinear dependence between variables. Its value is zero when the variables are independent. Dependency graphs are constructed using the Random Forest model to measure variable dependence, since tree-based models can inherently capture complex relationships. In the dependency graph approach, the coefficient of determination ($R^2$) of the fitted model is used as a quantitative measure of dependence, either between descriptor sets and the target property or among different descriptor sets. In both methods, the input descriptor sets are first reduced to relevant descriptors using the pipeline as described in Section \ref{rel_desc}. Detailed descriptions of both correlation analysis approaches are provided in Section \ref{corr_anaylsis}.

Figure \ref{fig:distance_corr} shows heatmaps of the distance correlation values obtained for ``max\_pfc" and ``Cv\_25" using the relevant descriptors from each descriptor set. In both cases, we observe that descriptors from the ``MATMINER" descriptor set, derived from structure and compositions, have a higher association with the target material properties than the ``LOBSTER" descriptor set, derived from quantum-chemical bonding descriptors. Note that the ``MATMINER" descriptor set includes many more descriptors that encode information about elements, the local geometric bonding environment, and structural symmetry. However, our descriptor set ``LOBSTER" also shows some independent association with the target property. For the case of ``max\_pfc", our descriptors have a stronger association than for ``Cv\_25". This aligns well with our results from the descriptor relevance rankings, where we saw our descriptors achieve the top ranking for the ``max\_pfc" target. Also, on combining the descriptor sets, i.e., ``MATMINER+LOBSTER", we see an increase (increment beyond the standard deviation) in the distance correlation value only for ``max\_pfc". For the other targets, the distance‑correlation values of our descriptors fall within the same range as those of the ``MATMINER" set alone (see the associated GitHub pages of our repository). This suggests that, while relevant, our descriptors are unlikely to add substantial additional information when machine learning these properties. Based on these results, we expect that the combined descriptor set ``MATMINER+LOBSTER" will yield the largest improvement in predictive performance for the ``max\_pfc" target, with minimal to no improvement on the other targets.

The dependency graph analysis shows similar trends (available on the GitHub pages of our repository). It is also worthwhile to note that our results on dependency graphs demonstrate that it is possible to learn some of the relevant quantum-chemical bonding descriptors using the ``MATMINER" descriptor sets with reasonably good predictive accuracy ($R^2 >$  0.9) for all the targets considered in this work, even with simple tree-based models such as Random Forests. For instance, the ICOHP corresponding to the strongest bond, ionicity, distance at the strongest bond can be learned. A detailed examination of which of the ``MATMINER" descriptors were most important for learning the bonding descriptors will not be presented here, as it falls beyond the scope of this work.

These findings suggest a promising pathway toward the development of intermediate, descriptor-oriented graph neural networks, as proposed in Ref.~\cite{gouvea2026combining}, that can be computationally efficient while capturing chemically meaningful bonding information. In such an approach, surrogate models could be trained to directly predict quantum-chemical bonding descriptors from crystal structures, thereby bypassing the need for explicit LOBSTER calculations. As shown in the following section, incorporating explicitly calculated quantum-chemical bonding descriptors into the models improves predictive performance for some target properties.

\subsection{Model training and evaluation}

\begin{table}[h]
\centering
\caption{Overview of Random Forest Regressor (RF) and MODNet model performance from our 5-fold cross-validation (CV) tests using both feature sets for various material properties. Reported metrics are Mean Absolute Error (MAE $\pm \text{ std}$). std denotes the standard deviation observed across MAE in five folds. The `*' besides the target property indicates that a significant improvement in model performance is observed based on the corrected resampling t-test conducted using 10-fold CV results, with the following mapping: * $p<0.05$, ** $p<0.01$, and *** $p<0.001$.}
\begin{tabular}{llcccccccc}
\toprule
\multirow{1}{*}{\textbf{Target}} &
%\multirow{1}{*}{\textbf{Metric}} &
\multicolumn{2}{c}{\textbf{RF}} &
\multicolumn{2}{c}{\textbf{MODNet}} \\
\cmidrule(lr){2-3} \cmidrule(lr){4-5}
 & \textbf{MATMINER} & \textbf{MATMINER+LOBSTER} & \textbf{MATMINER} & \textbf{MATMINER+LOBSTER} \\
\midrule
\multirow{1}{*}{last\_ph\_peak [$\mathrm{cm^-1}$]} 
 & 39.509 $\pm$ 4.915 & 39.198 $\pm$ 4.110 & 33.348 $\pm$ 7.193 & 30.639 $\pm$ 3.409\\
\midrule
\multirow{1}{*}{max\_pfc [eV/\AA$^2$]*} 
 & 1.582 $\pm$ 0.177 & 1.385 $\pm$ 0.186 & 1.243 $\pm$ 0.184 & 1.057 $\pm$ 0.188 \\
\midrule
\multirow{1}{*}{log\_g\_vrh***} 
 & 0.096 $\pm$ 0.004 & 0.091 $\pm$ 0.006 & 0.067 $\pm$ 0.004 & 0.066 $\pm$ 0.004 \\
\midrule
\multirow{1}{*}{log\_k\_vrh**} 
 & 0.085 $\pm$ 0.006 & 0.079 $\pm$ 0.005 & 0.055 $\pm$ 0.005 & 0.053 $\pm$ 0.005 \\
\midrule
\multirow{1}{*}{log\_klat\_300**} 
 & 0.190 $\pm$ 0.003 &  0.185 $\pm$ 0.002 & 0.159 $\pm$ 0.006 & 0.154 $\pm$ 0.009 \\
\midrule
\multirow{1}{*}{log\_kp\_300**} 
 & 0.224 $\pm$ 0.001 & 0.216 $\pm$ 0.002 & 0.192 $\pm$ 0.008 & 0.185 $\pm$ 0.008 \\
\midrule
\multirow{1}{*}{log\_msd\_all\_300*} 
 & 0.076 $\pm$ 0.001 & 0.075 $\pm$ 0.001 & 0.059 $\pm$ 0.001 & 0.059 $\pm$ 0.003 \\
\midrule
\multirow{1}{*}{log\_msd\_all\_600*} 
 & 0.079 $\pm$ 0.001 & 0.078 $\pm$ 0.001 & 0.062 $\pm$ 0.001 & 0.059 $\pm$ 0.001 \\
\midrule
\multirow{1}{*}{log\_msd\_max\_300} 
 & 0.084 $\pm$ 0.004 & 0.083 $\pm$ 0.003 & 0.069 $\pm$ 0.004 & 0.071 $\pm$ 0.006 \\
\midrule
\multirow{1}{*}{log\_msd\_max\_600} 
 & 0.086 $\pm$ 0.003 & 0.086 $\pm$ 0.003 & 0.073 $\pm$ 0.003 & 0.073 $\pm$ 0.003 \\
\midrule
\multirow{1}{*}{log\_msd\_mean\_300} 
 & 0.069 $\pm$ 0.001 & 0.067 $\pm$ 0.002 & 0.051 $\pm$ 0.002 & 0.053 $\pm$ 0.004 \\
\midrule
\multirow{1}{*}{log\_msd\_mean\_600*} 
 & 0.072 $\pm$ 0.002 & 0.072 $\pm$ 0.002 & 0.053 $\pm$ 0.004 & 0.055 $\pm$ 0.004 \\
\bottomrule
\end{tabular}
\label{tab:model_comparison_rf_modnet}
\end{table}

\begin{figure}[ht]
\centering
\includegraphics[width=0.90\linewidth]{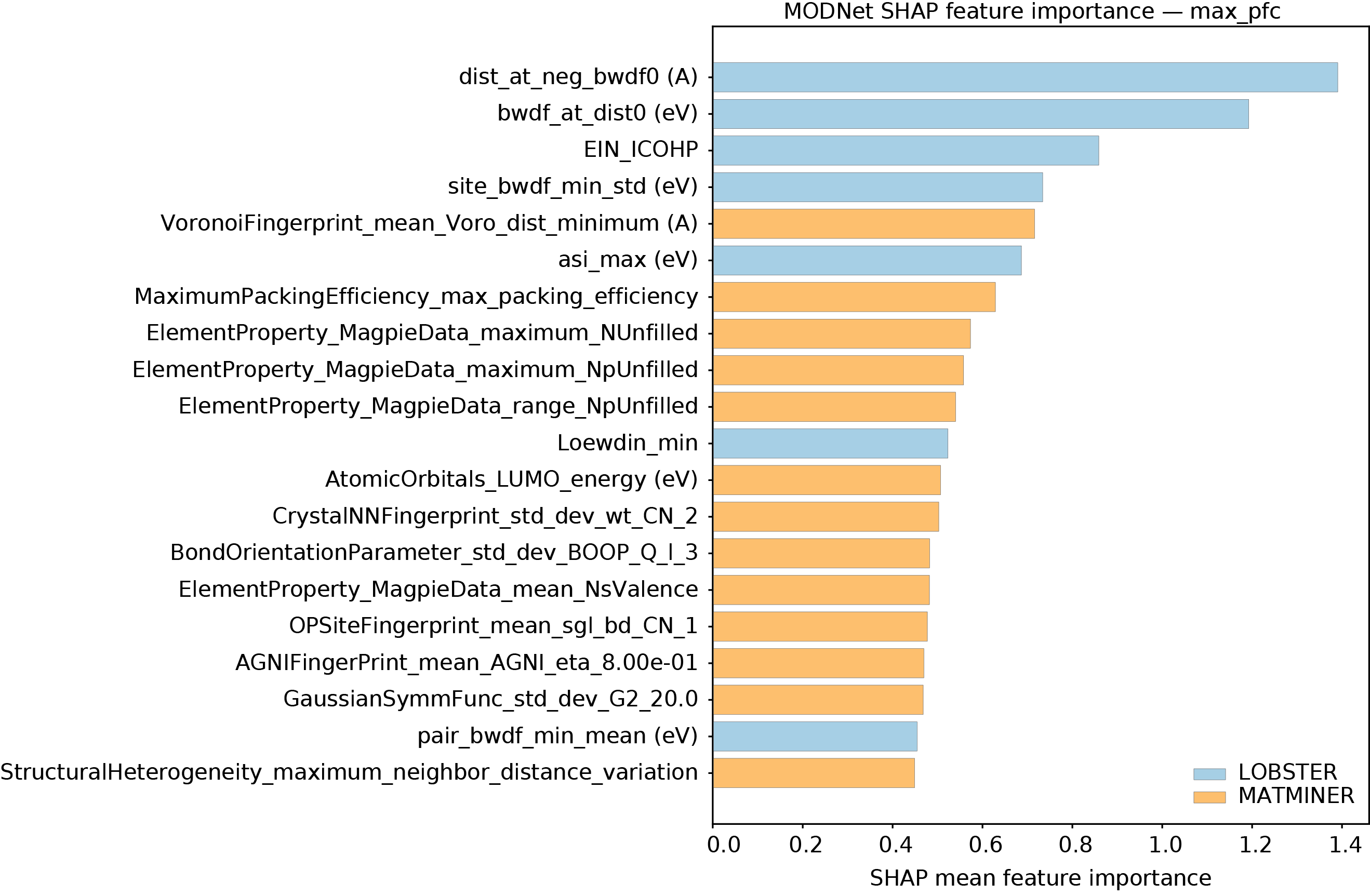}
\caption{Descriptor ranking as SHAP scores for the maximum of bond-projected force (max\_pfc) constant from MODNet model, where the legend distinguishes descriptor groups derived from ``MATMINER" (structure- and composition-based) and ``LOBSTER" (extracted from LOBSTER calculation data).}\label{fig:modnet_max_pfc_pfi_shap}
\end{figure}

To quantify the predictive capability of the descriptor sets, we train machine learning models for all targets. For model training, we use Random Forest (RF) and MODNet \cite{de2021materials, de2021robust} as machine learning model architectures, both of which are well-suited for the typical dataset sizes in our study. Table \ref{tab:model_comparison_rf_modnet} summarizes the results from the trained ML models' performance in 5-fold cross-validation (CV) tests. 
From the initial 5-fold CV results, we observe some reductions in mean absolute error (MAE) for some target properties when using the ``MATMINER+LOBSTER" descriptor set, specifically for ``last phonon dos peak (last\_ph\_peak)", ``max\_pfc", ``lattice thermal conductivity (log\_klat\_300 and log\_kp\_300)", ``bulk modulus (log\_k\_vrh)", ``shear modulus (log\_g\_vrh)" and ``mean squared displacement (log\_msd\_all\_300 and log\_msd\_all\_600)". To understand which descriptors drive these predictions, we conducted SHAP and permutation feature importance (PFI) analyses on the 5-fold CV models. These XAI methods capture complementary aspects of descriptor influence: SHAP quantifies each descriptor’s contribution to the prediction, whereas PFI measures the model's sensitivity to perturbations in descriptor values. Figure~\ref{fig:modnet_max_pfc_pfi_shap} shows the results of SHAP analysis for the MODNet model predicting max\_pfc. PFI identifies a similar set of influential descriptors (not shown here; see the GitHub pages of our repository), though rankings differ as expected due to the distinct evaluation metrics. In both analyses, quantum-chemical bonding descriptors are consistently more influential than structural or compositional features, which is consistent with the approximately linear correlation we observe between max\_pfc and the strongest bond ICOHP (Figure~\ref{fig:peak_pfc_icohp}b).

While the 5-fold CV results provide valuable preliminary insights, they cannot serve as a rigorous basis for comparing model performance. The training sets in different folds overlap substantially, violating the assumption of independent observations, and with only five folds, variance estimates are unstable and statistical power is low.\cite{nadeau1999inference, dietterich1998approximate} Consequently, apparent improvements may reflect sampling noise rather than genuine descriptor set advantages. Thus, we performed a corrected resampled paired t-test \cite{nadeau1999inference} on paired 10-fold CV results at a significance level of $\alpha = 0.05$ to quantify whether the observed performance difference is statistically significant (see Section \ref{model_training_method} for additional details). Improvements are considered significant if the p-value is below 0.05 for either the RF or MODNet model.

\begin{figure}[ht]
\centering
\includegraphics[width=0.95\linewidth]{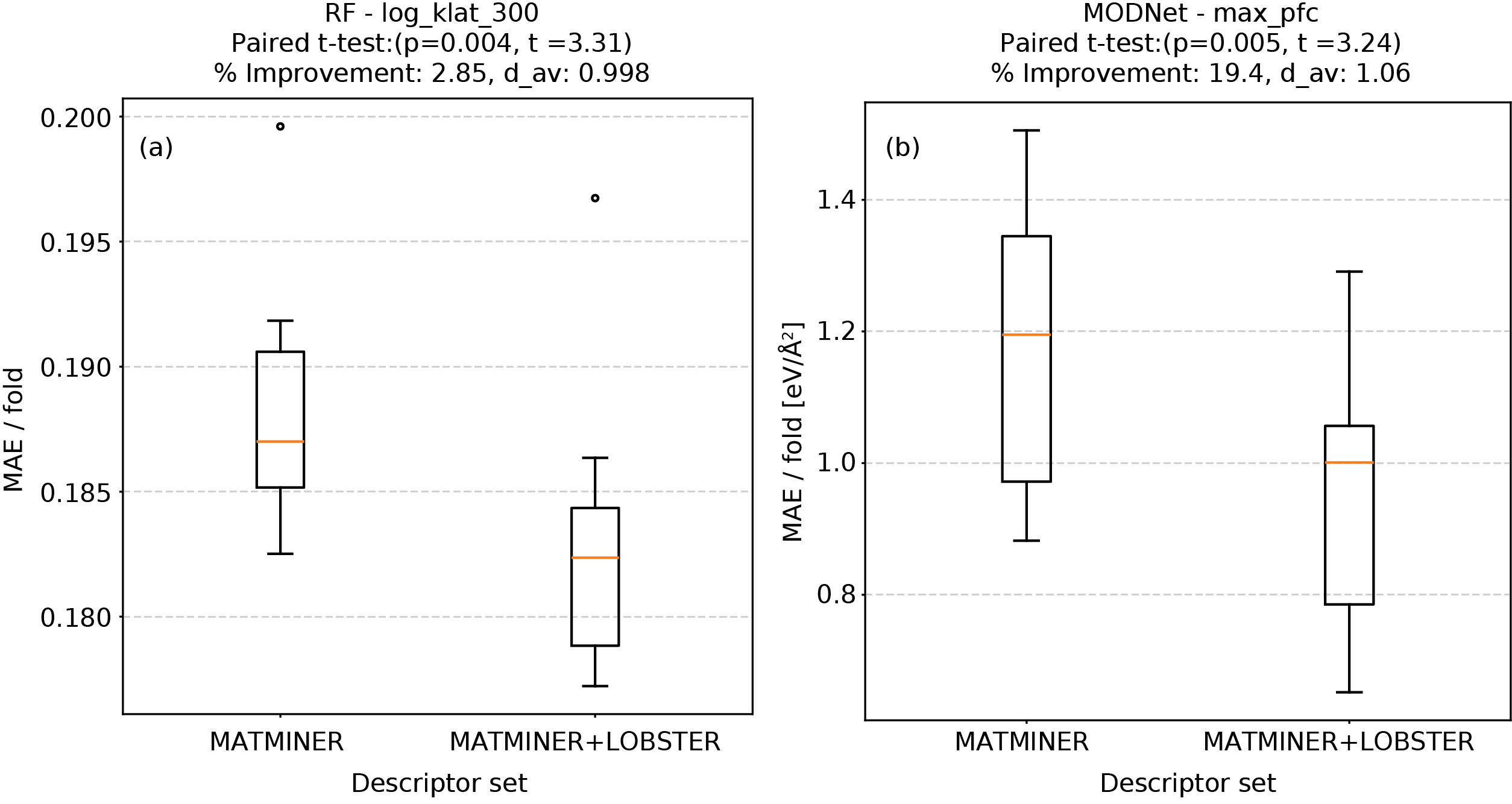}
\caption{Corrected resampling t-test based on per-fold mean absolute errors from 10-fold cross-validation for model comparison using the ``MATMINER" (structure and composition) and ``MATMINER+LOBSTER" (extracted from LOBSTER calculation data) descriptor sets. a) Total lattice thermal conductivity at 300 K, log\_klat\_300, from RF model shows a 2.85 \% improvement with the larger feature set, and b) Maximum of bond-projected force constant, max\_pfc, from MODNet model shows a 19.5 \% improvement.}\label{t_test}
\end{figure}

Based on these statistical tests, the ``MATMINER+LOBSTER" descriptor set leads to significant performance improvements for the following target properties: ``max\_pfc", ``lattice thermal conductivity (log\_klat\_300 and log\_kp\_300)", ``bulk modulus (log\_k\_vrh)", ``shear modulus (log\_g\_vrh)" and ``mean squared displacement (log\_msd\_all\_300 and log\_msd\_all\_600)". For the heat capacity ($C_\mathrm{v}$), internal energy ($U$), vibrational entropy ($S$), and Helmholtz free energy ($H$), although some of our descriptors appear in the list of top descriptors that drive model learning (SHAP and PFI analysis of 5-fold CV models),  there is no evident enhancement in the model's predictive performance relative to simply using site, structure, and composition-derived descriptors from Matminer, as also seen from our significance tests results. Table \ref{tab:model_comparison_rf_modnet_si} shows a 5-fold CV relative performance overview for RF and MODNet for Cv\_*, U\_*, H\_*, and S\_*. Table \ref{tab:t_test} summarizes the corrected resampling t-test results for all the targets. In Figure \ref{t_test}, we show the results from our corrected resampling t-test for ``log\_klat\_300" (left) and ``max\_pfc" (right) targets from the RF and MODNet models, respectively. 

We now briefly consider the interesting case of ``last\_ph\_peak". This property is strongly correlated with ``max\_pfc" as shown in Figure~\ref{fig:peak_pfc_icohp}. SHAP and permutation feature importance (PFI) analyses of the 5-fold CV models indicate that the same quantum-chemical bonding descriptors identified for max\_pfc also contribute to learning this target, although at lower relative importance. This observation involves the same descriptor highlighted in our earlier work on this property \cite{naik2023quantum} (ICOHP corresponding to the strongest bond), though its influence is less pronounced here. In terms of predictive performance, when comparing models trained with and without these descriptors using paired 10-fold CV results, their inclusion results in about 5\% reduction in mean MAE (see Table \ref{tab:t_test}) for the MODNet model; however, this improvement does not reach statistical significance under the corrected resampled t-test. These findings suggest that, while these quantum-chemical bonding descriptors still contribute meaningfully by reducing the standard deviation of MAE across folds, indicating improved model stability, their additional predictive benefit is not statistically distinguishable from that of a model trained with advanced structure-based descriptors.

The significance test results show that, in general, the more complex MODNet models achieve lower MAEs as expected, while RF models more often produce statistically significant improvements. This is because MODNet models sometimes exhibit greater fold-to-fold variability in MAEs, which inflates the standard error in the corrected resampled t-test and reduces its power, making improvements observed harder to be deemed significant. Figure~\ref{fig:modnet_struggle_example} in the Supplementary information illustrates one such example.

Lastly, for the properties showing significant improvements in predictive performance (see Table~\ref{tab:t_test}), we trained symbolic regression (SR) models with SISSO using the top 20 descriptors identified from our relevant descriptor selection step (i.e., results from Section~\ref{rel_des_results}). Figure~\ref{fig:arfs_feat_imp} illustrates an example of these descriptors for the SISSO model targeting the ``max\_pfc" property; descriptors for the other targets can be found on the GitHub Pages of our code repository. Recent studies have shown that including SR-derived descriptors in model training can further enhance predictive performance \cite{foppa2022hierarchical, shmuel2024symbolic, gouvea2026combining}, suggesting that these descriptors could also be incorporated in future work. Here, we present the results from one-dimensional SISSO models obtained at rung one for ``max\_pfc" and ``log\_klat\_300" on the full dataset. Rung one means that the mathematical operators are applied only once. The coefficients in these SISSO-derived expressions were determined via least squares regression.

\begin{align}
\mathrm{max\_pfc} &= 1.77 
    -4.25\left(\mathrm{\frac{ bwdf_{at\_dist0} }{ dist_{at\_neg\_bwdf0} } }\right)
\end{align}

\begin{align}
    \mathrm{log\_klat\_300} &= -0.41 
    -2.9\left(\mathrm{\frac{ pair_{bwdf\_skew\_mean} }{ DensityFeatures_{vpa} } }\right)
\end{align}

A closer examination shows that, for a subset of targets exhibiting performance improvements, the trained symbolic regression (SR) models also favor quantum-chemical bonding descriptors in their derived expressions. For example, in the case of the ``max\_pfc”, which measures bond stiffness in a material, the SISSO models preferentially construct analytic expressions composed of the strongest bond strength ($\mathrm{bwdf_{at\_dist0}}$) and its corresponding bond length ($\mathrm{dist_{at\_neg\_bwdf0}}$). This observation is consistent with prior studies that have demonstrated relationships between bond lengths \cite{bagheri2023identification, zahn2025experimental}, bond strengths \cite{deringer2015vibrational, hempelmann2022orbital}, and force constants. The simplest one-dimensional (number of expressions combined linearly by SISSO) expression identified by SISSO for ``max\_pfc", i.e., $\left(\mathrm{\frac{ bwdf_{at\_dist0} }{ dist_{at\_neg\_bwdf0} } }\right)$, represents the ratio of the strongest bond strength to its associated bond length and exhibits a strong negative correlation with the target (Pearson correlation coefficient: $-0.91$), consistent with the convention that stronger bonds correspond to more negative ICOHP values.

For lattice thermal conductivity at 300 K (log\_klat\_300), descriptors derived from quantum-chemical bonding information have a considerable influence on model predictions, as indicated by SHAP and permutation feature importance (PFI) rankings from 5-fold CV models (see the GitHub Pages of the code repository). These descriptors capture bonding heterogeneity in a material, including bond-weighted distribution function skewness ($\mathrm{pair\_bwdf\_skew}$), the asymmetry index (asi), and the effective interaction number ($\mathrm{EIN_{ICOHP}}$). This aligns with prior studies highlighting the role of bonding heterogeneity in reducing lattice thermal conductivity.\cite{pal2018bonding, sato2021bonding, jin2022bonding, powell2025exploiting} Notably, the simple, interpretable expressions constructed by SISSO also incorporate these descriptors. The one-dimensional SISSO expressions for log\_klat\_300 with rung 1 and 2 are: $(\mathrm{\frac{pair_{bwdf\_skew\_mean} }{ DensityFeatures\_vpa}})$ and $\left(\mathrm{\frac{ pair_{bwdf\_skew\_mean} }{ DensityFeatures\_vpa }} \right) * \left(\sqrt[3]{ \mathrm{EIN\_ICOHP }}\right)$, respectively. Both expressions are negatively correlated with log\_klat\_300 (Pearson correlation coefficients $-0.66$ and $-0.71$, respectively), indicating that increased bonding heterogeneity and larger volume per atom tend to reduce thermal conductivity. The second expression (Rung 2), with the stronger correlation of $-0.71$, alone accounts for approximately 52\% of the variance ($R^2$) across a diverse set of 3,200 materials. These correlations are higher than recently proposed normalized ICOHP (0.46) and ICOBI (0.47) descriptors when computed on the dataset used in this work (see Figure~\ref{fig:kappa_desc_si}).\cite{al2025accelerated} Consistent with the classical Slack model and a previous ML study using structural and compositional features, volume per atom is also a critical descriptor, exhibiting an inverse relationship with lattice thermal conductivity.\cite{morelli2006, wang2020identification} Overall, these findings corroborate prior observations that both bonding heterogeneity and larger volume per atom are among key factors in reducing lattice thermal conductivity. They further demonstrate that interpretable expressions obtained from SISSO can effectively capture these effects.

For the bulk modulus (log\_k\_vrh), shear modulus (log\_g\_vrh), and mean-squared displacements (log\_msd\_*), SHAP and permutation feature importance (PFI) scores (see the GitHub Pages of the code repository) from 5-fold CV models indicate that a similar set of quantum-chemical bonding descriptors, as observed for maximum of bond-projected force constants and lattice thermal conductivity, also contributes to predictions. This makes intuitive sense, as elastic moduli and atomic vibrations are governed by bond strengths and local bonding heterogeneity: stiffer, more uniform bonds lead to higher moduli and smaller mean-squared displacements. However, the simple, one-dimensional expressions constructed by SISSO do not explicitly include these descriptors. These observations can likely be attributed to two factors: (i) complex interactions among descriptors that SISSO’s simple, low-dimensional expressions cannot fully capture, and (ii) the unit constraints imposed during model construction, which may exclude certain descriptor combinations that remain important in more flexible models such as Random Forest or MODNet.

To further assess the influence of descriptor pre-screening on the derived SISSO expressions, we additionally employed a descriptor selection based on a genetic algorithm (GA). Thus, we train SISSO models on an optimal subset of 20 descriptors selected via the GA approach. We start the GA approach after our initial descriptor selection step (i.e., the results from Section~\ref{rel_des_results}). While the resulting symbolic expressions are not always identical, both selection approaches consistently identify bonding-related descriptors among the key contributors for the same set of targets, as we report above using the top twenty descriptors approach. The discrepancy between the expressions obtained from both approaches is not surprising, as symbolic regression solutions are not inherently unique, and diverse descriptor subsets can capture the underlying relationship with comparable accuracy through distinct functional forms. Nevertheless, our results confirm that the physically meaningful signal captured by these descriptors is robust to the selection procedure. Full results of the GA-assisted SISSO models are available in the Zenodo repository.

\section{Methods}\label{methods}

\subsection{Descriptors}

The conventional descriptors derived from structures and compositions are generated using the \textit{Matminer2023Featurizer} preset from the MODNet package \cite{de2021materials, de2021robust}, which serves as a convenient wrapper providing access to nearly all implementations available in the Matminer package. This featurizer comprises a diverse set of structure- and composition-based features that have demonstrated strong predictive performance in previous studies using MODNet models. To summarize, these descriptors consist of statistics such as mean, median, standard deviation, maximum, and minimum, generated from elemental properties or geometric information available from the input structure.  It also includes descriptors used to accelerate ab initio simulations, such as atom-centered Gaussian symmetry functions\cite{behler2011atom} and AGNIFingerprints,\cite{botu2015adaptive, khorshidi2016amp} that aim to provide a numerical representation of the sites' local environments. Details regarding each descriptor obtained are not elaborated on here; interested readers are asked to refer to the original publications and the documentation of the Matminer package for further information.\cite{ward2018matminer} Everywhere in the text, descriptors from this group are referred to as ``MATMINER".

The quantum-chemical bonding descriptors (referred to as ``LOBSTER") are generated using the featurizer module of the LobsterPy package that was extended as part of this study.\cite{george2022automated, naik2024lobsterpy} These descriptors, inspired by the conventional descriptors from structure, consist of statistics such as mean, standard deviation, minimum, and maximum extracted by analyzing the most relevant bonds. Details on the algorithm for identifying the most relevant bonds in a structure can be found in our initial publication on automation.\cite{george2022automated} In addition to these, we also extract statistical descriptors from bond-weighted distribution functions (BWDFs) and bonding asymmetry index (ASI), first introduced by Deringer et al. \cite{deringer2014bonding} and Belli et al. \cite{belli2026chemical}, respectively. Conceptually, ASI is similar to the centrosymmetry parameter\cite{kelchner1998dislocation} or the local inversion‑symmetry‑breaking order parameter ($F_{\mathrm{IS}}$),\cite{milkus2016local} but uses bond strength to quantify the local asymmetry. The computation of BWDFs and ASIs has been implemented in LobsterPy as part of this work. One can also extract such descriptors using COOPs (ICOOPs), COHPs (ICOHPs), or COBIs (ICOBIs). In this study, however, we have limited the evaluation to descriptors extracted using COHPs, ICOHPs and atomic charges(Mulliken/L{\"o}wdin). Please note that, when training models to predict mean-squared displacement parameters for symmetrically inequivalent sites within the structure, the descriptors used within our ML models are associated with these specific sites. Formulations and detailed descriptions of these descriptors are provided in the Supplementary information Section \ref{desc_formulations} and Table \ref{tab:bonding_descriptors}, respectively.

\subsection{Datasets}

Characterization of the structural and compositional diversity of all target material properties datasets reported below are available from our code repository \href{https://digimatchem.github.io/paper-ml-with-lobster-descriptors/}{GitHub pages}. Similar to our bonding analysis dataset, the target material properties datasets also comprise sufficiently diverse classes of materials.

\subsubsection{Vibrational and Thermodynamic Properties}

The vibrational and thermodynamic properties for 1520 inorganic compounds are retrieved from the prior work by Petretto et al.\cite{petretto2018high}. The last phonon density of states peak (last\_ph\_peak) extracted from this dataset (1264 datapoints) is among the benchmark targets for evaluating the accuracy of machine learning models in the Matbench leaderboard.\cite{dunn2020benchmarking}. In one of our previous works, together with some of the authors of this work, we demonstrated that ICOHPs (associated with the strongest bond) exhibit a strong correlation with this target property, and that model performance was enhanced by including this descriptor.\cite{naik2023quantum} The authors would like to emphasize that in prior work, a minimal set of descriptors, predominantly comprising structural and element-based descriptors, was used. In our current work, we aim to predict the last phonon peak again, but with a significantly larger and more advanced set of descriptors that also capture the local environments of the structure based on geometric information.

We also extracted the maximum of the bond-projected force constants (max\_pfc) for the same set of 1264 datapoints.  The bond-projected force constant (a measure of bond stiffness) is defined as \cite{deringer2014bonding} 
\begin{equation}
\phi_B = \Phi_{ij}\hat{d_{ij}} =\left|\left(\begin{array}{lll} 
\phi_{i j}^{x x} & \phi_{i j}^{x y} & \phi_j^{x z} \\
\phi_{i j}^{y x} & \phi_{i j}^{y y} & \phi_{i j}^{y z} \\
\phi_{i j}^{z x} & \phi_{i j}^{z y} & \phi_{i j}^{z z}
\end{array}\right)\left(\frac{\vec{r}_j-\vec{r}_i}{\left|\vec{r}_j-\vec{r}_i\right|}\right) \right|\label{eq:pfc}
\end{equation}

Here,  $\Phi_{ij}$ and $\hat{d_{ij}}$ denote the force constant matrix and unit bond vector for atoms \textit{i} and \textit{j} in a structure, respectively. This property is interesting because phonons are typically derived from underlying force constants, and one would expect the bond-projected force constant (pFC) to exhibit some relationship with vibrational frequencies (last\_ph\_peak) and bond strengths. There have been prior works focused on a particular set of materials, where the relationship between pFC and quantum-chemical bonding descriptors, namely ICOHPs, has been reported.\cite{friak2018ab, deringer2014bonding, hempelmann2021long} Li et al. \cite{li2018theoretical} have linked the pFCs with weak bonding, leading to higher anharmonicity and lower lattice thermal conductivity. Whereas Bagheri et al. \cite{bagheri2023identification} have used pFC to screen for two-dimensional materials. However, to the authors' knowledge, no large-scale studies have demonstrated a relationship between the bonding descriptors, pFC, and phonon peak. The Figure \ref{fig:peak_pfc_icohp} below illustrates the relationship between the ``last phonon peak” and ``max\_pfc” (left panel), as well as the correlation between the strongest bond ``ICOHP” and ``max\_pfc" (right panel) within these materials.

\begin{figure}[ht]
    \centering
    \includegraphics[width=0.95\linewidth]{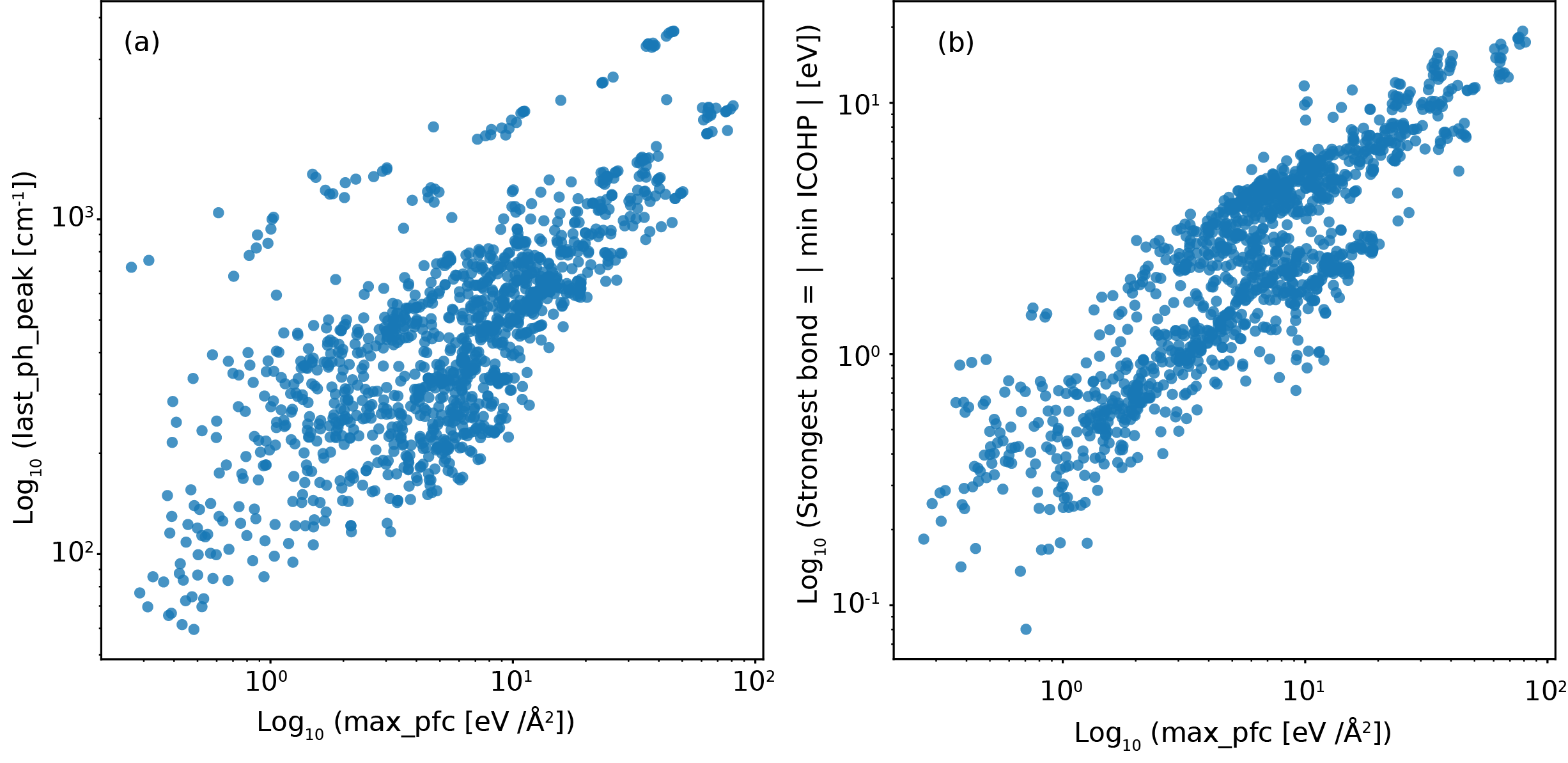}
    \caption{Parity plots showing correlation between (a) Last peak of phonon density of states ,``last\_ph\_peak” and Maximum of bond-projected force constant, ``max\_pfc” (b) Absolute value of the strongest bond, i.e, minimum ``ICOHP” in a material and ``max\_pfc".}
    \label{fig:peak_pfc_icohp}
\end{figure}

To compute the thermodynamic properties, we use only database entries labeled as dynamically stable (i.e., free of imaginary modes), yielding a total of 1,245 data points. Imaginary modes indicate a breakdown of the harmonic approximation, so that these entries are excluded as the definition of the thermodynamic properties would not hold. For these dynamically stable entries, we extract vibrational entropy ($S$), Helmholtz free energy ($H$), internal energy ($U$), and heat capacity ($C_\mathrm{v}$) at 25, 305, and 705 K. For the same set of entries, we calculate thermal displacement parameters, including mean-squared displacements (MSDs). For extracting MSDs, convergence tests were performed using mesh sizes from 40 to 220 \AA~(increments of 20 \AA) and frequency cutoffs of 0.0, 0.1, and 0.13 THz. Such frequency cutoffs have been used in previous work by George et al. \cite{george2015anisotropic} to compute thermal displacement parameters. A frequency cutoff of 0.1 THz was found to stabilize the convergence of the computed MSDs, and mesh sizes of approximately 100–140 \AA~were generally sufficient. MSDs for a given site were considered converged if the difference was less than 0.001 \AA$^2$ across three consecutive mesh-size increments. For materials achieving convergence, MSDs of symmetrically inequivalent sites were extracted, and the log${10}$ values of these site-specific MSDs were used as targets. This procedure results in a total of 3,691 data points at 300 K and 600 K. Additionally, the log${10}$ of the mean and maximum MSD values across sites were used as separate targets at 300 K and 600 K, yielding 1,157 data points for each.

\subsubsection{Lattice Thermal conductivity}

The average Peierls (kp) and total lattice thermal conductivities (klat) at 300 K are obtained from the partially published anharmonic phonon properties database for approximately 6000 inorganic materials.\cite{ohnishi2025database} Out of these structures, we have bonding descriptors computed for 3299 structures. After excluding unphysical and low calculation accuracy entries using the filtering criteria as employed by authors of the publication (exclude entries with a large phonon gap ($\ge$ 10) and kp $\ge$ 500, excessive kp, i.e., $\ge$ 2000 and fitting errors in force constants of order two and three larger than 10 \%) we have in total of 3249 data points for this target. Log$10$ values
of ``klat" and ``kp" are used as targets.

\subsubsection{Elasticity}
The elasticity dataset is curated by retrieving elasticity data from the Materials Project database \newline(v2025.09.25),\cite{jain2013materials} which were computed using the methodology described in Ref. \cite{de2015charting}. Cleaning of this dataset is achieved by employing the same criteria as described in the matbench dataset.\cite{dunn2020benchmarking} Thus, we exclude entries having formation energy (or energy above the convex hull) more than 150 meV and those having negative $G_\mathrm{{Voigt}}$, $G_\mathrm{{Reuss}}$, $G_\mathrm{{VRH}}$, $K_\mathrm{{Voigt}}$, $K_\mathrm{{Reuss}}$, or  $K_\mathrm{{VRH}}$ and those failing $G_\mathrm{{Reuss}} \leq G_\mathrm{{VRH}} \leq G_\mathrm{{Voigt}}$ or  $K_\mathrm{{Reuss}} \leq K_\mathrm{{VRH}} \leq K_\mathrm{{Voigt}}$ and those containing noble gases. Furthermore, we remove entries labeled as metals from this filtered set, resulting in 3,079 datapoints for bulk (k\_vrh) and shear modulus (g\_vrh), respectively, as the minimal basis set that we employ here for the projection in LOBSTER might not be sufficient for metals.

\subsection{Evaluation of descriptors}\label{descriptor_eval}

\subsubsection{Identifying relevant descriptors}\label{rel_desc}

Our initial descriptor space contains over 2,500+ columns, not all of which are expected to contribute to model learning. We first remove nearly constant and highly correlated descriptors to eliminate uninformative or redundant descriptors. Next, we apply an all-relevant feature selection (ARFS) method \cite{kursa2010feature, Bury2025ThomasBury}, which identifies all descriptors that contribute to learning the target property (labeled as ``relevant" by the feature selection method) and may still include potentially redundant ones. This avoids the risk that the new bonding‑based descriptors are prematurely discarded by minimal‑subset feature‑selection methods. Only in this way can we properly assess their predictive utility. The descriptors labeled as ``relevant" by this process are then used as inputs to the machine learning models.
Additional details on the entire process are provided in Section~\ref{rel_desc_details} of the Supplementary information. Figure~\ref{fig:mlpipeline} illustrates the overall model training and evaluation workflow.

\begin{figure}[ht]
    \centering
    \includegraphics[width=0.85\linewidth]{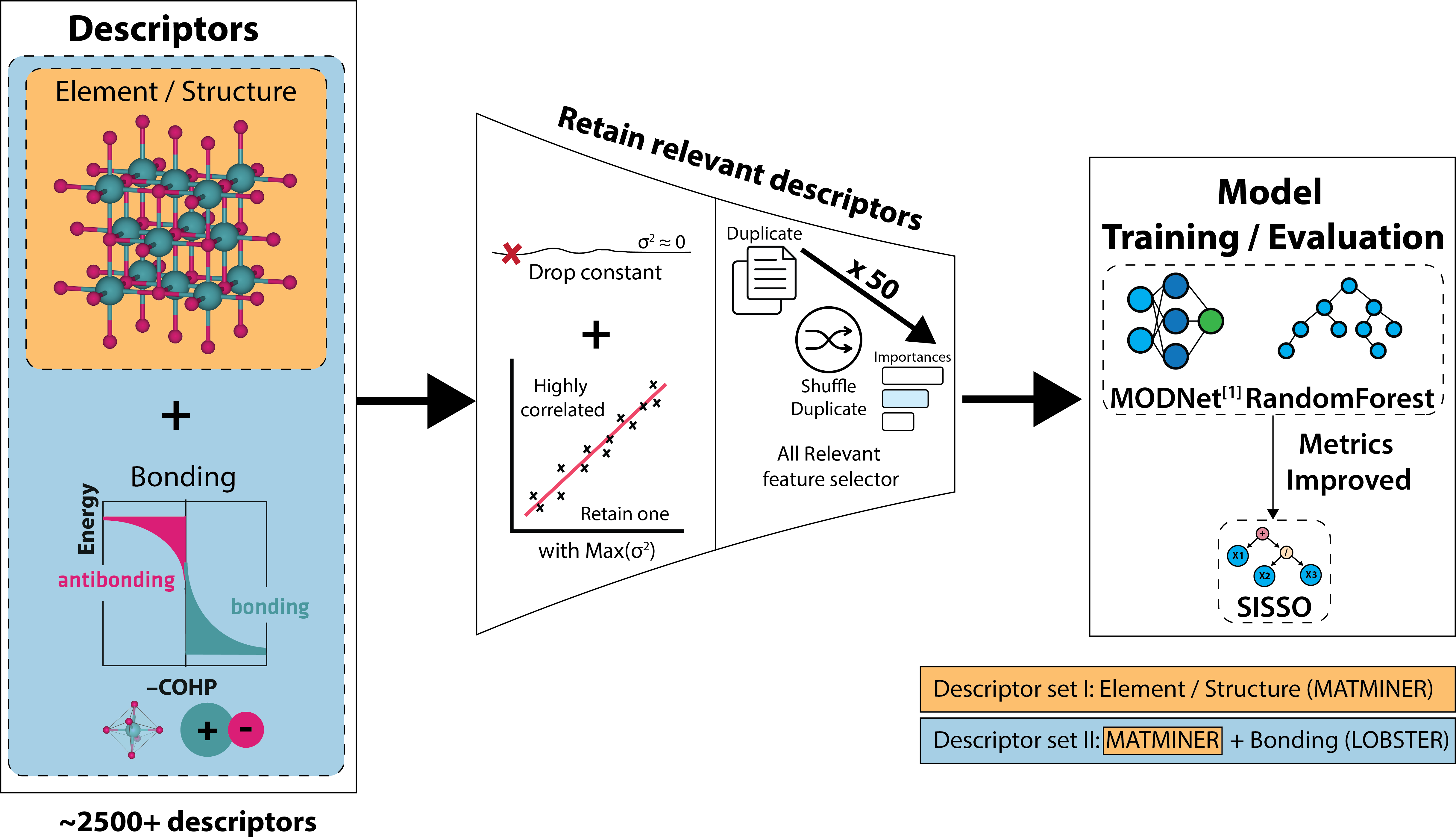}
    \caption{Schematic depicting the model training pipeline. Both models, RF and MODNet, are trained and evaluated using both descriptor sets.}
    \label{fig:mlpipeline}
\end{figure}

\subsubsection{Descriptor set and target correlation analysis}\label{corr_anaylsis}

Prior to correlation analysis, irrelevant descriptors are removed using the feature selection pipeline described in Section~\ref{rel_desc}. This procedure is applied separately to the ``MATMINER", ``LOBSTER", and ``MATMINER+LOBSTER" descriptor sets, and the resulting relevant descriptors are used as inputs for the analyses described below.

In the first approach, we compute distance correlation and distance covariance metrics, which capture both linear and nonlinear dependencies between variables and are well suited for multivariate data \cite{ramos_carreno_torrecilla_2023_dcor, ramos_carreno_2022_7484447, das2024feature}. A distance correlation value of zero indicates statistical independence between two variables, whereas a value of one corresponds to perfect dependence. Distance covariance is used in conjunction with distance correlation to perform statistical tests of independence. For each target property and descriptor set, distance correlation values are estimated via bootstrap sampling (500 samples with replacement) and repeated 20 times to ensure robustness. Statistical tests of independence are performed using distance covariance, with small $p$-values (i.e., $p<0.05$) indicating rejection of the null hypothesis of independence. The averaged distance correlation values from the bootstrap runs are visualized as heatmaps. Representative example can be found in Section~\ref{corr_analysis_results}.

In the second approach, a dependency graph is constructed using Random Forest (RF) models trained under a fivefold cross-validation (5-fold CV) scheme. A multi-output RF regressor is used to quantify dependencies among descriptor sets, while univariate RF regressors assess the relationship between each descriptor set and the target property. Edge weights in the dependency graph correspond to the mean $R^2$ values across folds. In addition, we evaluate the learnability of individual ``LOBSTER" descriptors from ``MATMINER" descriptors using univariate RF regression, with average $R^2$ values reported as bar charts. Results from the dependency graph analysis are available on the project’s GitHub repository webpage.

\subsubsection{Model training and evaluation} \label{model_training_method}

To assess the influence of descriptor choice on machine-learned model performance, we train Random Forest (RF) and EnsembleMODNet (MODNet) models using two descriptor spaces: one consisting solely of structure- and composition-based descriptors (``MATMINER’’) and another augmented with bonding-based descriptors (``MATMINER+LOBSTER’’). For each model architecture, two corresponding models are trained. Model training follows a 5-fold cross-validation (CV) protocol, where the feature selection pipeline described in Section~\ref{rel_desc} is applied exclusively to the training folds, and performance is evaluated on the held-out test folds. RF models are trained using 500 estimators without further hyperparameter tuning. For MODNet models, the built-in genetic algorithm-based hyperparameter optimization implemented in the MODNet package is applied only to the training folds. The genetic algorithm-based hyperparameter optimization implementation of MODNet also optimizes for feature scaling besides other model hyperparameters. Further implementation details can be found in ref.~\cite{de2022accurate} Both models are optimized using mean absolute error (MAE) as the objective function, and standard regression metrics, including $R^2$ and root mean square error (RMSE), are additionally evaluated. In the main text, we report the mean MAE and its standard deviation across CV folds; additional metrics are available in the associated GitHub repository. To interpret model predictions, we further apply explainable artificial intelligence (XAI) techniques, specifically SHAP and permutation feature importance (PFI), to all trained models.

To statistically evaluate whether the inclusion of quantum-chemical bonding-based descriptors improves predictive performance, we conduct paired 10-fold CV runs for both the RF and MODNet models. Identical train–test splits are enforced across descriptor sets by fixing the random seed, yielding paired fold-level test metrics. Statistical significance is assessed using the corrected resampled t-test proposed by Nadeau and Bengio \cite{nadeau1999inference}, which accounts for fold dependence. Fold-level MAE values are used, with per-fold differences defined as $d_i = \mathrm{MAE_{\text{MATMINER},i}} - \mathrm{MAE_{\text{MATMINER+LOBSTER},i}}$. A one-tailed test with $\alpha = 0.05$ is employed to test whether the augmented descriptor set yields lower MAE values. Because $p$-values ($p$) do not quantify effect magnitude and are sensitive to sample size, we additionally report effect size estimates.\cite{sullivan2012using,wasserstein2016asa,demvsar2006statistical} Specifically, we compute two measures using fold-level MAE values: Cohen’s $d_{\text{av}}$ which quantifies the magnitude of MAE improvement when adding LOBSTER descriptors relative to variability across cross-validation folds \cite{lakens2013calculating}, and the mean relative reduction in MAE (Mean RR), which expresses the same improvement as a percentage decrease in average prediction error. These quantities are defined as
%Two measures are computed using fold-level MAE values: Cohen’s $d_{av}$ \cite{lakens2013calculating} and the mean relative reduction in MAE, defined as
\begin{equation}
    d_{\text{av}} = \mathrm{\frac{\mu_{MATMINER} - \mu_{MATMINER+LOBSTER}}{\frac{\sigma_{MATMINER} + \sigma_{MATMINER+LOBSTER}}{2}}}\label{eq:d_av}
\end{equation}

\begin{equation}
   \text{Mean RR} = \mathrm{\frac{\mu_{MATMINER} - \mu_{MATMINER+LOBSTER}}{\mu_{MATMINER}} * 100}\label{eq:mean_rr}
\end{equation}

\noindent where $\mu$ and $\sigma$ denote the mean and standard deviation of MAE values across the 10 folds.

If the statistical analysis of paired 10-fold CV results for RF or MODNet models indicates a significant improvement ($p < 0.05$) upon inclusion of the bonding-based descriptor set (``MATMINER+LOBSTER") relative to the baseline (``MATMINER"), we conclude that the bonding descriptors aid in learning the corresponding material property. For these properties, we subsequently train SISSO models on the full dataset to assess whether the bonding descriptors are also selected in simple, interpretable symbolic regression (SR) expressions. The output of SISSO is a compact mathematical equation, and the objective here is to explore whether intuitive relationships between bonding descriptors and the target property can be identified in their simplest functional form. Thus, the operator set and model complexity used in this work for training SISSO models are restricted as described further. Training SISSO models with large descriptor spaces is computationally expensive, especially when constructing higher-dimensional, more complex models. Several approaches have therefore been proposed to mitigate this cost, including descriptor pre-screening using Random Forests \cite{jiang2024boostingsissoperformancesmall}, genetic algorithms (GA) \cite{mazheika2024combining,shmuel2024symbolic}, or mutual information-based methods \cite{xu2022sisso}. In this work, we use the reduced descriptor set obtained from our feature reduction pipeline (Section~\ref{rel_desc}) and train SISSO models using the top twenty descriptors from this set. The mathematical operators used in SISSO are summarized in Equation~\ref{eq:sisso_operators}, where $\mathrm{\phi_1}$ and $\mathrm{\phi_2}$ denote descriptors in the primary descriptor vector $\Phi_0$. The superscript $(m)$ enforces physical unit consistency, such that operations between descriptors with incompatible units are excluded.

\begin{equation}
    \hat{H}^{(\mathrm{m})} \equiv\left\{+,-, \times, \div, \exp ,-\exp,\log,\vert \vert,\vert \phi_1 - \phi_2 \vert, \sqrt{},\sqrt[3]{},^{-1}, ^{2},^{3}\right\}\left[\phi_1, \phi_2\right],
    \label{eq:sisso_operators}
\end{equation}

For all SISSO models, the number of candidate descriptors retained in the sure-independence screening step ($n_{\text{sis}}$) is set to 100, the number of residuals ($n_{\text{residuals}}$) is set to 10, the maximum descriptor dimension is set to 3, and the maximum rung is set to 1. The resulting symbolic expressions are presented in the main text, while equations obtained at rung 2 are provided in the associated GitHub repository.

It is important to note that any form of descriptor pre-screening introduces a certain degree of bias into the resulting symbolic SR expressions. Our approach of selecting only the top 20 descriptors from the reduced set is subject to this same constraint. Thus, as an additional validation step, we also explored GA-assisted descriptor selection applied to the reduced descriptor set (Section~\ref{rel_desc}). Here, a genetic algorithm is used to identify an optimal subset of twenty descriptors that minimizes MAE, while keeping all SISSO hyperparameters fixed. 
%The GA implementation is different from the one presented in \cite{mazheika2024combining} and is based on binary differential evolution \cite{bde} using the DEAP package \cite{deap1,deap2}, and is conducted within a 5-fold CV framework.
The GA implementation differs from that presented in \cite{mazheika2024combining}; it is instead based on binary differential evolution \cite{bde}, implemented using the DEAP package \cite{deap1,deap2}, and conducted within a 5-fold cross-validation framework. In each generation, a population of 20 individuals is initialized by randomly selecting 20 descriptors from the reduced set, with the descriptor creation rung set to 2, maximum descriptor dimension set to 3, $n_{\text{sis}} = 100$, and $n_{\text{residuals}} = 10$. At the end of the GA search, the selected descriptor subset is used to train SISSO on the full dataset. We find that the GA-assisted approach produces descriptor sets that are broadly similar to those obtained by directly selecting the top 20 descriptors from the reduced set. Our current implementation of the GA-assisted strategy is slow in contrast to the feature selection pipeline discussed above, with long run times due to serial population evaluation. Additionally, given the vast combinatorial search space, the small population size of 20 per generation may limit the algorithm’s ability to fully explore possible solutions. Consequently, the efficient application of this approach may require improvements, such as parallel population evaluation. Nevertheless, representative results obtained using this approach for selected target properties are made available via our Zenodo repository, and the code implementation is provided in the associated GitHub repository.

\section{Conclusion}

We systematically evaluated quantum-chemical bonding descriptors for predicting bonding-related material properties and assessed their relevance, complementarity with conventional descriptors, and impact on machine learning performance. Note that our evaluation is based on diverse datasets spanning the continuum from covalent to ionic bonding. This ensures that our findings are not limited to a specific material class but are broadly applicable. Our analyses show that these descriptors consistently carry predictive information across all 11 material properties, although their influence varies across targets. To contextualize these results, it is worth noting that existing featurizer implementations in the Matminer package served as inspiration for the design of these bonding descriptors. Many Matminer featurizers, developed over the years, also encode bonding information from structural geometry and were included in the descriptor space for this study. 

 A key insight is that our descriptors significantly improve predictions for directional or local properties, such as the maximum of bond-projected force constant, bulk and shear moduli, lattice thermal conductivity, and mean-squared displacements, likely because these properties are strongly influenced by bond strengths and local bonding heterogeneity. Predictions of the maximum bond-projected force constant show the largest improvement, with MAE reduced by approximately 19\%. In contrast, global, averaged thermodynamic properties, such as heat capacity or vibrational entropy, show little benefit from these descriptors, as they depend more on overall structural averages than on directional bonding details.

Perhaps most notably, symbolic regression using SISSO uncovers interpretable structure–property relationships based on quantum‑chemical bonding descriptors, providing insight into the underlying chemical relationships. For max\_pfc, the ratio of the strongest bond (minimum ICOHP) to its associated bond length shows a strong linear correlation with the target, with a Pearson correlation coefficient of $r =-0.91$. For lattice thermal conductivity at 300 K (log\_klat\_300), a descriptor combining bonding heterogeneity and the average number of interactions per atom, normalized by the volume per atom, correlates negatively with the target ($r =-0.71$). Both expressions align with the trends reported in prior studies for these properties.\cite{deringer2015vibrational,pal2018bonding,hempelmann2021long,sato2021bonding} These expressions are physically intuitive and also offer a computationally inexpensive route for preliminary materials screening.

Overall, our results open several avenues for further exploration. Firstly, ICOBI-based descriptors could complement ICOHPs: unlike ICOHPs, ICOBIs are inherently normalized within a fixed range, which may improve the stability of models sensitive to feature scaling. Their strong correlation with ICOHP, however, may limit predictive gains and thus needs to be quantified in further studies. Secondly, bond-informed local environment descriptors, such as coordination environments, could potentially be improved by using ICOHPs or ICOBIs to guide connectivity, so that only chemically significant bonds are included. This approach may yield chemically meaningful structural descriptors, potentially improving model performance. Thirdly, unsupervised analysis of energy-resolved COHP/COBI or projected density-of-states (PDOS) data from LOBSTER could reveal clusters of materials with similar bonding or electronic structure. Such clustering could be used to prioritize materials for further investigation or to guide targeted screening for specific properties. Finally, in the context of graph neural networks (GNNs), integrating bonding information as edge attributes alongside conventional bond distances can provide additional chemical information. Additionally, optimizing the number of edges in input graphs using ICOHP or ICOBI thresholds can yield sparser graphs, potentially accelerating GNN training; such computational speed-ups are expected to be significant for large datasets, although the effects on predictive performance should be carefully evaluated.

%Other promising directions include unsupervised learning on energy-resolved COHP/COBI or projected density-of-states data from LOBSTER, and integrating bonding information as edge attributes into graph neural networks (GNN) training, alongside conventional bond distances. Alternatively, optimizing the number of edges in input graphs using ICOHPs or ICOBIs could produce sparser graphs, potentially accelerating GNN training. Such speed-ups are likely to be noticeable primarily in large datasets.

% Acknowledgements
\medskip
\section*{Acknowledgements} \par %delete if not applicable))
J.G. and A.A.N. were supported by ERC Grant MultiBonds (grant agreement no. 101161771; funded by the European Union. Views and opinions expressed are however those of the author(s) only and do not necessarily reflect those of the European Union or the European Research Council Executive Agency. Neither the European Union nor the granting authority can be held responsible for them.) G.-M.R. is Research Director at the Fonds de la Recherche Scientifique (FNRS). We gratefully acknowledge the Gauss Centre for Supercomputing e.V. (www.gausscentre.eu) for providing generous computing time on the GCS Supercomputer SuperMUC-NG at the Leibniz Supercomputing Centre (www.lrz.de) under Project No. pn73da. We thank Dr. Jan Hempelmann for granting access to his implementation of the bond-projected force-constant code and for helpful feedback on the manuscript. We are grateful to Prof. Dr. Purcell for promptly resolving the issues with the SISSO++ code encountered during this work. We thank Prof. Dr. Luca Ghiringhelli for sharing the GA-SISSO implementation, for discussions that inspired our GA-based feature selector for SISSO, and for feedback on the manuscript. We also thank Dr. Rog\'{e}rio Almeida Gouv\^{e}a for providing constructive feedback on the manuscript. ChatGPT has been used to refine the language. The whole manuscript content has been carefully checked by the authors.

\section*{Data availability statement}
All the trained models and post-processed data used to generate the figures and tables are available for download from Zenodo.\cite{naik_2026_18329982} All data and code used to reproduce the ML results of this work are available in the GitHub repository associated with this work.\cite{naik_2026_18599928}

\medskip
\newpage
\makeatletter
\renewcommand*{\theHsection}{SI.\arabic{section}}
\renewcommand*{\theHsubsection}{SI.\arabic{section}.\arabic{subsection}}
\makeatother
\setcounter{figure}{0} % reset figure counter for Sup. Figures
\setcounter{equation}{0} % reset equation counter for Sup. Equations
\setcounter{table}{0} % reset equation counter for Sup. Equations
\setcounter{page}{1} % reset page count
\setcounter{section}{0}
\renewcommand{\thesection}{S\arabic{section}} 
\renewcommand{\thesubsection}{\thesection.\arabic{subsection}}
\renewcommand{\thefigure}{S\arabic{figure}}
\setcounter{figure}{0}
\makeatletter
\renewcommand{\thetable}{S\@arabic\c@table} % make Figure legend start with Fig. S.
\makeatother
\makeatletter
\renewcommand{\theequation}{S\@arabic\c@equation} % make Figure legend start with Fig. S.
\makeatother

% \phantomsection
% \section*{Supplementary Information For}
% \addcontentsline{toc}{section}{Supplementary Information}
% \begin{center}
%     \textbf{A critical assessment of bonding descriptors for predicting materials properties}
% \end{center}

% --- SI title ---
\phantomsection
\addcontentsline{toc}{section}{Supplementary Information} % adds SI to TOC

\begin{center}
    {\LARGE \textbf{Supplementary Information}}\\[1em] % main SI title
    {\large \textbf{A critical assessment of bonding descriptors for predicting materials properties}} % paper title
\end{center}

\noindent\author{Aakash Ashok Naik}
\author{Nidal Dhamrait}
\author{Katharina Ueltzen}
\author{Christina Ertural}
\author{Philipp Benner}
\author{Gian-Marco Rignanese}
\author{Janine George*}

\begin{affiliations}
\noindent Aakash Ashok Naik, Katharina Ueltzen, Prof. Dr. Janine George \\
Address \\
Department of Materials Chemistry, Federal Institute for Materials Research and Testing, Berlin, 12205, Germany \\
Institute of Condensed Matter Theory and Optics, Friedrich Schiller University Jena, Jena, 10587, Germany \\
Email Address: janine.george@bam.de \\
\hfill \break
\noindent Nidal Dhamrait, Dr. Christina Ertural \\ % affiliation of the time when the contributions were made
Address \\
Department of Materials Chemistry, Federal Institute for Materials Research and Testing, Berlin, 12205, Germany \\
\hfill \break
\noindent Dr. Philipp Benner \\
Address \\
eScience Group, Federal Institute for Materials Research and Testing, Berlin, 12205, Germany \\
\hfill \break
\noindent Prof. Dr. Gian-Marco Rignanese \\
Address \\
Institute of Condensed Matter and Nanosciences (IMCN), UCLouvain, Louvain-la-Neuve, 1348, Belgium \\

\end{affiliations}

\section{Details: Identifying relevant descriptors}\label{rel_desc_details}

The relevant descriptor identification consists of three main steps. First, nearly constant features—those that remain unchanged for more than 95\% of the training dataset—are removed as uninformative. Second, highly correlated features (Pearson correlation coefficient $>$ 0.9) are grouped, and only one feature from each group is retained based on variance. Third, an all-relevant feature selection (ARFS, GrootCV implementation \cite{Bury2025ThomasBury}) is applied to identify all descriptors that contribute to learning the target property, including potentially redundant ones. In this step, shadow descriptors are generated by randomly perturbing the original descriptors and added to the dataset. An ensemble LightGBM model \cite{ke2017lightgbm} is then trained on the extended dataset, and descriptor importance is computed using SHAP values. Descriptors whose importance falls below the maximum of the median importance scores of the shadow descriptors are labeled irrelevant. The process is repeated across multiple iterations using K-fold cross-validation, with the number of iterations serving as an iteration parameter that determines how many shadow descriptor sets are generated and evaluated. A higher number of iterations increases the robustness of estimates of descriptor importance. Convergence tests indicate that fifty iterations produce a stable set of relevant descriptors, as measured by the Jaccard similarity index \cite{jaccard1912distribution} approaching or exceeding 0.9. Stability is considered achieved when the Jaccard similarity between successive iterations reaches or exceeds this threshold. This implementation ensures that selected descriptors are robust, reproducible, and reflective of their contribution to model learning.

\section{Descriptor Formulations}\label{desc_formulations}

\phantomsection
\subsection*{Average interaction strength and effective interaction number}
\addcontentsline{toc}{subsection}{Average interaction strength and effective interaction number}

The quantities outlined below were originally introduced by Peter M{\"u}ller in his doctoral thesis,\cite{muller2019quantum} and the notations used are as per the reference number \cite{nelson2023141}. The weight of a specific bond, for instance, $i$ in a structure, is determined by its bond strength ($\mathrm{ICOHP}_i$) relative to the total bond strength ($\mathrm{ICOHP}_{total}$). Utilizing these weights ($w_i$), one can compute the weighted COHP ($\overline{\mathrm{COHP}}(E)$)  and weighted ICOHP ($\overline{\mathrm{ICOHP}}$) as shown in equations \ref{eq:S2} and \ref{eq:S3}, respectively. Besides, this one can also compute the effective interaction number per atom within a structure, which is solely based on bond strengths and is comparable to a coordination number, as depicted in equation \ref{eq:S4}, where $N_{atoms}$ denotes the total number of atoms in the structure.

\begin{align}
    w_i &=\frac{\mathrm{ICOHP}_i}{\mathrm{ICOHP}_{\text {total}}}\label{eq:S1} \\
    \overline{\mathrm{COHP}}(E) &=\sum_i\left(w_i \cdot \operatorname{COHP}_i(E)\right)\label{eq:S2} \\ 
    \overline{\mathrm{ICOHP}}&=\int_{-\infty}^{\varepsilon_{\mathrm{F}}} \overline{\mathrm{COHP}}(E) \mathrm{d}E\label{eq:S3} \\
    \mathrm{EIN} &= \frac{\mathrm{ICOHP}_{\text {total }}}{\overline{\mathrm{ICOHP}}} \cdot \frac{2}{N_{\text {Atoms }}}\label{eq:S4}
\end{align}

\phantomsection
\subsection*{COXX shape based descriptors}
\addcontentsline{toc}{subsection}{COXX shape based descriptors}

The $n^{th}$ moment from a COXX curve is computed using the following formulation,

\begin{align}
 n^{th}_{moment} &= \frac{\int (p^n \cdot \text{COXXs} \, dE)}{\int \text{COXXs} \, dE}\label{eq:S5} 
\end{align}

\noindent where $p$ denotes the energy (E). Here, \textbf{COXX} generically represents energy-resolved bonding indicators and may correspond to \textbf{COOP}, \textbf{COHP}, or \textbf{COBI}, depending on the input data used. The $1^{st}$, $2^{nd}$, $3^{rd}$ and $4^{th}$ moment are named as center\_COXX, width\_COXX, skewness\_COXX, kurtosis\_COXX, respectively. In the present work, only \textbf{COHP}-based descriptors are extracted and evaluated. 

\phantomsection
\subsection*{Charge based descriptors}
\addcontentsline{toc}{subsection}{Charge based descriptors}

To compute ionicity of a structure using Mulliken or L{\"o}wdin atomic charges following formulation as introduced in reference \cite{nelson2023141} is used. Here $q_i$ and $v_{eff, i}$ denote the atomic charge and effective valence for atom $i$ in a structure, respectively.

\begin{align}
 \mathrm{Ionicity}_{\text{Charges}} &= \frac{1}{N_{\text {Atoms }}} \sum_i^{N_{\text {Atoms }}}\left(\frac{q_i}{v_{\text {eff }, i}}\right)\label{eq:S6}
\end{align}

\noindent Besides Ionicity, simple statistical information based on atomic charges (Mulliken and L{\"o}wdin), such as minimum, maximum, mean, and standard deviation, is also computed for each structure.

\phantomsection
\subsection*{Bond weighted distribution functions}
\addcontentsline{toc}{subsection}{Bond weighted distribution functions}

The bond-weighted distribution function (BWDF) is essentially a histogram of bond strengths as a function of interatomic distance. This concept and formulation were introduced by Deringer et al. \cite{deringer2014bonding} in the LOBSTER program. The bonding indicator can be ICOHPs, ICOBIs, or ICOOPs. Functionality to compute this quantity for the entire structure (bwdf\_\*), for each unique atom pair in the structure (pair\_bwdf\_*), per site (site\_bwdf\_*), or per bond label, is implemented in LobsterPy \cite{naik2024lobsterpy} package as part of this work, thereby bypassing the need to explicitly rerun the LOBSTER program to obtain this quantity.

\begin{align}
\mathrm{BWDF}&=\sum_{\mathrm{B}>\mathrm{A}}\left[\delta\left(r-\left|\mathbf{r}_{\mathrm{AB}}\right|\right) \times B_{\mathrm{AB}}\right]\label{eq:S7}
\end{align}

In the formula above, $\delta$ denotes the Dirac delta function, $r$ represents the interatomic distance between atoms A and B, and $B_{\mathrm{AB}}$ signifies the bonding indicator between atoms A and B. 

\phantomsection
\subsection*{Quantifying symmetry (asymmetry) of the atom's local bonding environment}
\addcontentsline{toc}{subsection}{Quantifying symmetry (asymmetry) of the atom's local bonding environment}

The centrosymmetry parameter~\cite{kelchner1998dislocation} and the local inversion‑symmetry‑breaking order parameter ($F_{\mathrm{IS}}$)\cite{milkus2016local} are widely used to quantify local symmetry in solids, specifically solids with defects and amorphous materials. The centrosymmetry parameter compares pairs of opposite-neighbor vectors, while $F_{\mathrm{IS}}$ measures deviations from local inversion symmetry via force imbalances, providing a complementary perspective on structural disorder.
Similar to these concepts, Belli et al. \cite{belli2026chemical} introduced the asymmetry index (ASI), which quantifies local symmetry by combining bonding strength and directionality. The asymmetry index (ASI) for an individual atom $x$ is defined as:

\begin{align}
 \mathrm{ASI}_{x} &= \frac{1}{B_{x}} \sum_{\alpha=1}^{B_{x}} \mathrm{iCOXX}(x, \alpha) \, \boldsymbol{i}_{x \alpha}\label{eq:S8}   
\end{align}

\noindent Here, \textbf{iCOXX (x, $\alpha$)} represents the bonding indicator (ICOHP, ICOBI, or ICOOP) for the interaction between the atom \textbf{\textit{x}} and its neighbor \textbf{\textit{$\alpha$}}, and \textbf{$i_{x \alpha}$} is the unit vector indicating the direction of this interaction. Summation is performed over all neighboring atoms within the user-defined threshold (maximum distance). $B_x$ denotes the total number of interactions included in the sum.

\vspace{1em}
\noindent In this work, we consider ICOHP-based ASI statistics to quantify the degree of local bonding asymmetry within a structure. Values of the ASI close to zero indicate a highly symmetrical local environment, whereas higher values indicate greater asymmetry. The resulting $ASI_x$ is a vector for each atom, which is then converted to a scalar value by computing its Euclidean norm ($\left\lVert \mathrm{ASI}_{x} \right\rVert$). This scalar quantifies the overall asymmetry of the local bonding environment for atom $x$ in a structure. Finally, statistics like minimum, maximum, mean and standard deviation are used as descriptors. 

\begin{table}[ht]
\centering
\caption{List of bonding descriptors and corresponding descriptions}
\renewcommand{\arraystretch}{1.25}
\begin{tabular}{|l|p{10.5cm}|}
\hline
\textbf{Descriptor Group} & \textbf{Description} \\ \hline

\multicolumn{2}{|l|}{\textbf{ICOHP based}} \\ \hline

Icohp\_mean\_* & Mean ICOHP statistics (avg, max, min, std) per bond at symmetrically inequivalent sites. \\ \hline

Icohp\_sum\_* & Summed ICOHP statistics (avg, max, min, std) per bond at symmetrically inequivalent sites. \\ \hline

w\_ICOHP & Weighted ICOHP computed from COHPs using corresponding ICOHPs as weights and effective interaction number (EIN) for the structure \\ \hline

EIN\_ICOHP & Effective interaction number (EIN) for the structure computed using w\_ICOHP \\ \hline

asi\_* \\
(*: sum, mean, std, min, max) 
& Measures the asymmetry of the local bonding environment using ICOHP as weights; high ASI indicates more distorted or asymmetric coordination. \\ \hline

pair\_bwdf\_*, site\_bwdf\_*, bwdf\_* \\ 
(*: mean, min, max, std, skew, kurtosis)
& Statistics of bond-weighted distribution functions (BWDF) computed across atom pairs, sites, and overall structure. \\ \hline

bwdf\_at\_dist*
& sorted non-zero BWDF values. \\ \hline

dist\_at\_neg\_bwdf*
& Bond distances at which BWDF is non-zero. \\ \hline

bwdf\_0.0-* to bwdf\_*-5.0
& BWDF values for specific distance intervals. \\ \hline

\multicolumn{2}{|l|}{\textbf{COHP based}} \\ \hline

bonding\_perc* \\
(*: avg, max, min, std) & Statistics of the fraction of bonding contributions below the Fermi level from COHPs. \\ \hline

antibonding\_perc* \\
(*: avg, max, min, std) & Statistics of the fraction of anti-bonding contributions below the Fermi level from COHPs. \\ \hline

COHP\_*\\ (*: center, width, skewness, kurtosis, edge)
& Shape descriptors of the average COHP curve in the specified energy interval below the Fermi level: position, broadening, asymmetry, tail, and maximum value. \\ \hline

bnd\_wICOHP, antibnd\_wICOHP
& Percentage of bonding and antibonding interaction below the Fermi level computed from weighted COHPs, where ICOHPs are used as weights. \\ \hline

\multicolumn{2}{|l|}{\textbf{Charge }} \\ \hline

Madelung\_* (*: mull, loew) & Madelung energies computed from Mulliken (mull) and L{\"o}wdin (loew) charges. \\ \hline

Ionicity\_* (*: mull, loew) & Ionicity estimated from Mulliken (mull) and L{\"o}wdin (loew) charges. \\ \hline

Mulliken\_*  \\
(*: mean, max, min, std) & Mulliken Charge statistics from all sites. \\ \hline

Loewdin\_*  \\
(*: mean, max, min, std) & L{\"o}wdin Charge statistics from all sites. \\ \hline

\end{tabular}
\label{tab:bonding_descriptors}
\end{table}

\begin{table}[!ht]
\centering
\caption{Overview of Random Forest Regressor (RF) and MODNet model performance from our nested 5-fold CV tests using both feature sets for material properties, where bonding descriptors have almost no effect. Reported metrics are Mean Absolute Error (MAE $\pm \text{std}$). std denotes the standard deviation observed across MAE in five folds. }
\begin{tabular}{llcccccccc}
\toprule
\multirow{1}{*}{\textbf{Target}} &
\multicolumn{2}{c}{\textbf{RF}} &
\multicolumn{2}{c}{\textbf{MODNet}} \\
\cmidrule(lr){2-3} \cmidrule(lr){4-5}
 & \textbf{MATMINER} & \textbf{MATMINER+LOBSTER} & \textbf{MATMINER} & \textbf{MATMINER+LOBSTER} \\
\midrule
\multirow{1}{*}{$Cv_{25}$ [meV/atom]} 
 & 0.005 $\pm$ 0.000 & 0.005 $\pm$ 0.000 & 0.004 $\pm$ 0.000 & 0.004 $\pm$ 0.000 \\
\midrule
\multirow{1}{*}{$Cv_{305}$ [meV/atom]} 
 & 0.003 $\pm$ 0.000 & 0.003 $\pm$ 0.000 & 0.002 $\pm$ 0.000 & 0.003 $\pm$ 0.000 \\
\midrule
\multirow{1}{*}{$Cv_{705}$ [meV/atom]} 
 & 0.001 $\pm$ 0.000 & 0.001 $\pm$ 0.000 & 0.001 $\pm$ 0.000 & 0.001 $\pm$ 0.000 \\
\midrule
\multirow{1}{*}{$H_{25}$ [meV/atom]} 
 & 2.912 $\pm$ 0.287 & 2.932 $\pm$ 0.325 & 1.507 $\pm$ 0.193 & 1.505 $\pm$ 0.208 \\
\midrule
\multirow{1}{*}{$H_{305}$ [meV/atom]} 
 & 6.342 $\pm$ 0.407 & 6.371 $\pm$ 0.389 & 3.282 $\pm$ 0.331 & 3.298 $\pm$ 0.350 \\
\midrule
\multirow{1}{*}{$H_{705}$ [meV/atom]} 
 & 13.418 $\pm$ 0.795 & 13.447 $\pm$ 0.859 & 7.226 $\pm$ 0.719 & 7.258 $\pm$ 0.436 \\
\midrule
\multirow{1}{*}{$S_{25}$ [meV/atom]} 
 & 0.003 $\pm$ 0.000 & 0.003 $\pm$ 0.000 & 0.003 $\pm$ 0.000 & 0.003 $\pm$ 0.000 \\
\midrule
\multirow{1}{*}{$S_{305}$ [meV/atom]} 
 & 0.017 $\pm$ 0.001 & 0.017 $\pm$ 0.001 & 0.011 $\pm$ 0.001 & 0.010 $\pm$ 0.001 \\
\midrule
\multirow{1}{*}{$S_{705}$ [meV/atom]} 
 & 0.018 $\pm$ 0.001 & 0.018 $\pm$ 0.001 & 0.011 $\pm$ 0.001 & 0.011 $\pm$ 0.001 \\
\midrule
\multirow{1}{*}{$U_{25}$ [meV/atom]} 
 & 2.865 $\pm$ 0.312 & 2.882 $\pm$ 0.332 & 1.500 $\pm$ 0.190 & 1.485 $\pm$ 0.191 \\
\midrule
\multirow{1}{*}{$U_{305}$ [meV/atom]} 
 & 1.458 $\pm$ 0.393 & 1.480 $\pm$ 0.398 & 0.888 $\pm$ 0.099 & 0.896 $\pm$ 0.179 \\
\midrule
\multirow{1}{*}{$U_{705}$ [meV/atom]} 
 & 0.771 $\pm$ 0.207 & 0.763 $\pm$ 0.205 & 0.635 $\pm$ 0.093 & 0.631 $\pm$ 0.166 \\
\bottomrule
\end{tabular}
\label{tab:model_comparison_rf_modnet_si}
\end{table}

\begin{table}[ht]
\centering
\caption{Overview of corrected resampling t-test conducted on 10-fold CV results using Random Forest Regressor (RF) and MODNet model for comparing model performance difference is significant. Also includes effect size estimates, i.e., Cohen's $d_\mathrm{{av}}$, Relative MAE, and \% Folds Improvement measures.}
\resizebox{0.95\textwidth}{!}{%
\begin{tabular}{lccccccccccc}
\toprule
\thead{Target} & \thead{p-value \\ (RF)} & \thead{Significance \\ (RF)} & \thead{d\_av \\ (RF)} & \thead{Relative \\ MAE Improvement (RF)} & \thead{\% Folds \\ Improvement (RF)} &  & \thead{p-value \\ (MODNet)} & \thead{Significance \\ (MODNet)} & \thead{d\_av \\ (MODNet)} & \thead{Relative \\ MAE Improvement (MODNet)} & \thead{\% Folds \\ Improvement (MODNet)} \\
\midrule
last\_phdos\_peak & 0.452 &  & 0.015 & 0.318 & 50 & | & 0.166 &  & 0.268 & 5.406 & 70 \\
\midrule
\rowcolor{lobcolor} \textbf{max\_pfc} & 0.051 &  & 0.810 & 13.347 & 80 & | & 0.005 & ** & 1.064 & 19.446 & 100 \\
\midrule
\rowcolor{lobcolor} \textbf{log\_g\_vrh} & 0.000 & *** & 0.584 & 4.317 & 100 & | & 0.168 &  & 0.212 & 2.516 & 80 \\
\midrule
\rowcolor{lobcolor} \textbf{log\_k\_vrh} & 0.002 & ** & 0.661 & 6.997 & 100 & | & 0.134 &  & 0.185 & 2.405 & 80 \\
\midrule
\rowcolor{lobcolor} \textbf{log\_klat\_300} & 0.005 & ** & 0.998 & 2.850 & 90 & | & 0.051 &  & 0.612 & 3.610 & 80 \\
\midrule
\rowcolor{lobcolor} \textbf{log\_kp\_300} & 0.004 & ** & 0.881 & 3.150 & 100 & | & 0.083 &  & 0.576 & 2.567 & 70 \\
\midrule
\rowcolor{lobcolor} \textbf{log\_msd\_all\_300} & 0.018 & * & 0.361 & 1.489 & 80 & | & 0.055 &  & 0.715 & 3.856 & 90 \\
\midrule
\rowcolor{lobcolor} \textbf{log\_msd\_all\_600} & 0.017 & * & 0.356 & 1.473 & 90 & | & 0.067 &  & 0.551 & 3.044 & 80 \\
\midrule
log\_msd\_max\_300 & 0.719 &  & -0.061 & -0.738 & 50 & | & 0.508 &  & -0.003 & -0.045 & 60 \\
\midrule
log\_msd\_max\_600 & 0.745 &  & -0.068 & -0.795 & 50 & | & 0.893 &  & -0.219 & -3.157 & 10 \\
\midrule
log\_msd\_mean\_300 & 0.130 &  & 0.142 & 1.329 & 60 & | & 0.891 &  & -0.422 & -5.110 & 30 \\
\midrule
\rowcolor{lobcolor} \textbf{log\_msd\_mean\_600} & 0.022 & * & 0.193 & 1.912 & 90 & | & 0.449 &  & 0.023 & 0.271 & 70 \\
\midrule
Cv\_25 & 0.125 &  & 0.195 & 1.807 & 80 & | & 0.880 &  & -0.411 & -2.914 & 20 \\
\midrule
Cv\_305 & 0.531 &  & -0.008 & -0.101 & 30 & | & 0.604 &  & -0.074 & -1.418 & 50 \\
\midrule
Cv\_705 & 0.379 &  & 0.023 & 0.652 & 40 & | & 0.487 &  & 0.014 & 0.306 & 40 \\
\midrule
H\_25 & 0.390 &  & 0.023 & 0.291 & 50 & | & 0.348 &  & 0.053 & 1.196 & 40 \\
\midrule
H\_305 & 0.405 &  & 0.032 & 0.246 & 60 & | & 0.566 &  & -0.027 & -0.414 & 50 \\
\midrule
H\_705 & 0.801 &  & -0.153 & -1.144 & 40 & | & 0.030 & * & 0.325 & 3.108 & 90 \\
\midrule
S\_25 & 0.317 &  & 0.053 & 0.528 & 70 & | & 0.432 &  & 0.083 & 0.975 & 50 \\
\midrule
S\_305 & 0.714 &  & -0.077 & -0.621 & 30 & | & 0.339 &  & 0.171 & 1.789 & 40 \\
\midrule
S\_705 & 0.657 &  & -0.052 & -0.409 & 20 & | & 0.217 &  & 0.269 & 3.258 & 60 \\
\midrule
U\_25 & 0.660 &  & -0.027 & -0.357 & 50 & | & 0.677 &  & -0.052 & -1.082 & 30 \\
\midrule
U\_305 & 0.409 &  & 0.023 & 0.611 & 60 & | & 0.125 &  & 0.159 & 3.988 & 70 \\
\midrule
U\_705 & 0.803 &  & -0.079 & -2.173 & 40 & | & 0.320 &  & 0.196 & 4.924 & 70 \\
\bottomrule
\end{tabular}
}
\label{tab:t_test}
\end{table}

\begin{figure}[!ht]
\centering
\includegraphics[width=0.95\linewidth]{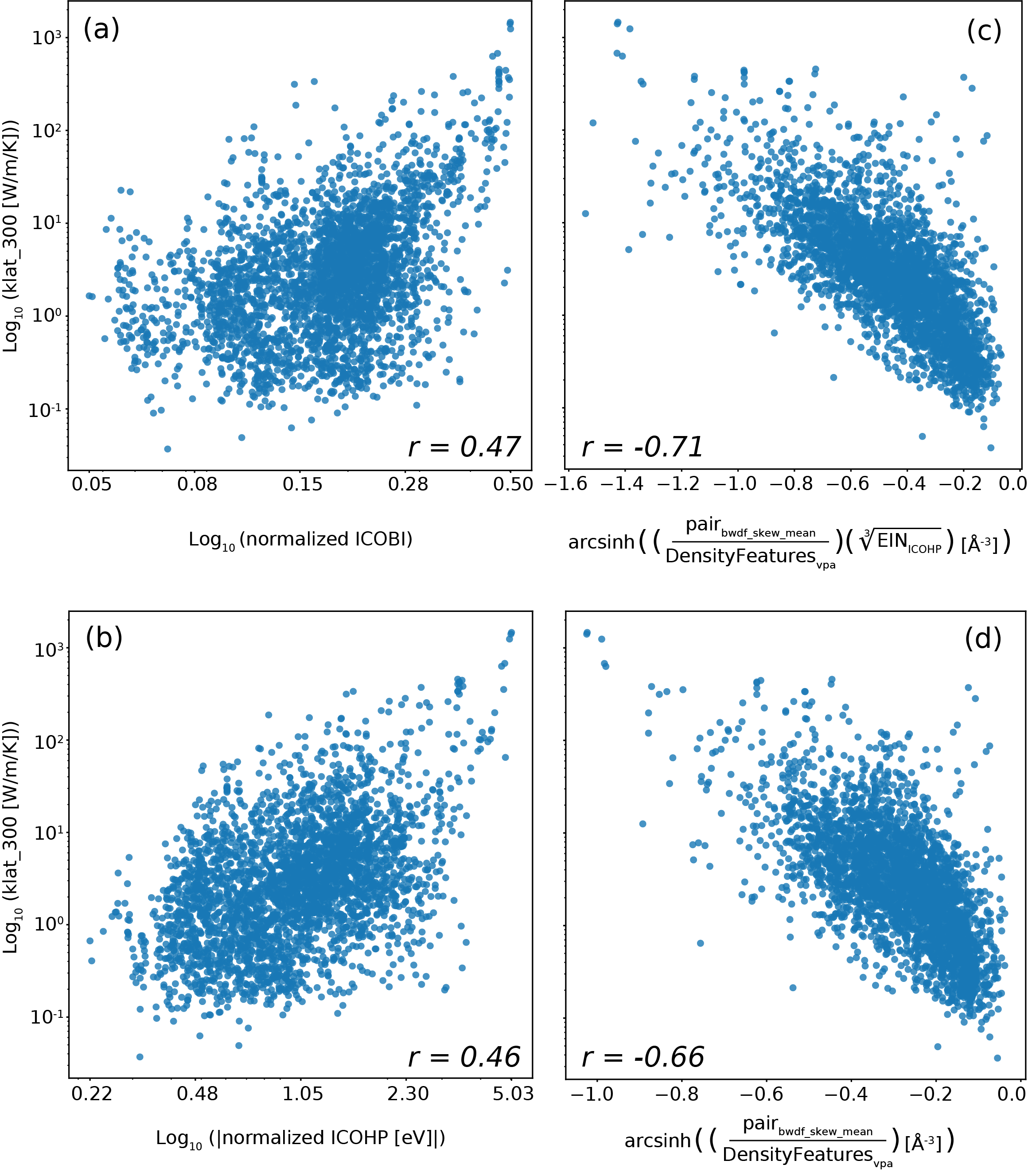}
\caption{Panels (a) and (b) show the correlation between $\log_{10}(\mathrm{klat\_300})$ and the normalized ICOBI and ICOHP descriptors, respectively, with the x-axis shown on a $\log_{10}$ scale. Panels (c) and (d) show the correlation between $\log_{10}(\mathrm{klat\_300})$ and the one-dimensional descriptors identified by SISSO in this work, which are composed of bonding descriptors. In these panels, the x-axis is scaled using the $\operatorname{arcsinh}$ transformation to preserve the sign of the correlation. $r$ denotes the Pearson correlation coefficient.}\label{fig:kappa_desc_si}
\end{figure}

\begin{figure}[!ht]
\centering
\includegraphics[width=0.80\linewidth]{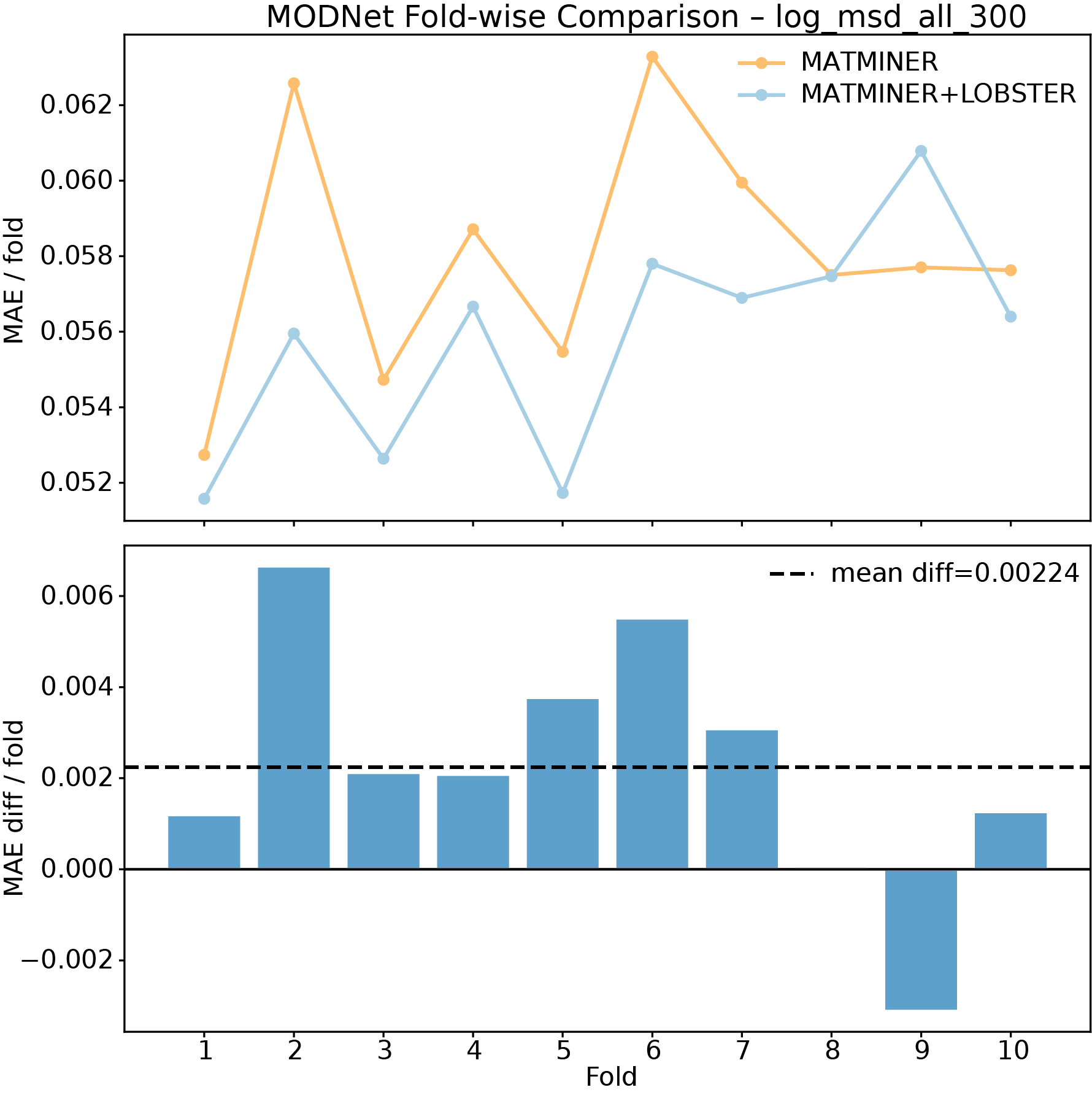}
\caption{Test-set MAE comparison of MODNet models from 10-fold cross-validation.
(Top) Per-fold MAE using ``MATMINER" and ``MATMINER+LOBSTER" descriptors. (Bottom) Per-fold MAE differences (``MATMINER" - ``MATMINER+LOBSTER"), showing high fold-to-fold variability that limits the reliability of significance testing.}\label{fig:modnet_struggle_example}
\end{figure}

\clearpage
\section{LOBSTER Bonding database overview}\label{database_overview}

The chemical composition categories are assigned using the most probable oxidation states as obtained with pymatgen, \cite{ong2013python} which are based on ICSD statistics. Periodic table heatmap and spacegroup distribution sunburst plots are generated using pymatviz.\cite{riebesell_pymatviz_2022}

\begin{figure}[!ht]
\centering
\includegraphics[width=0.8\linewidth]{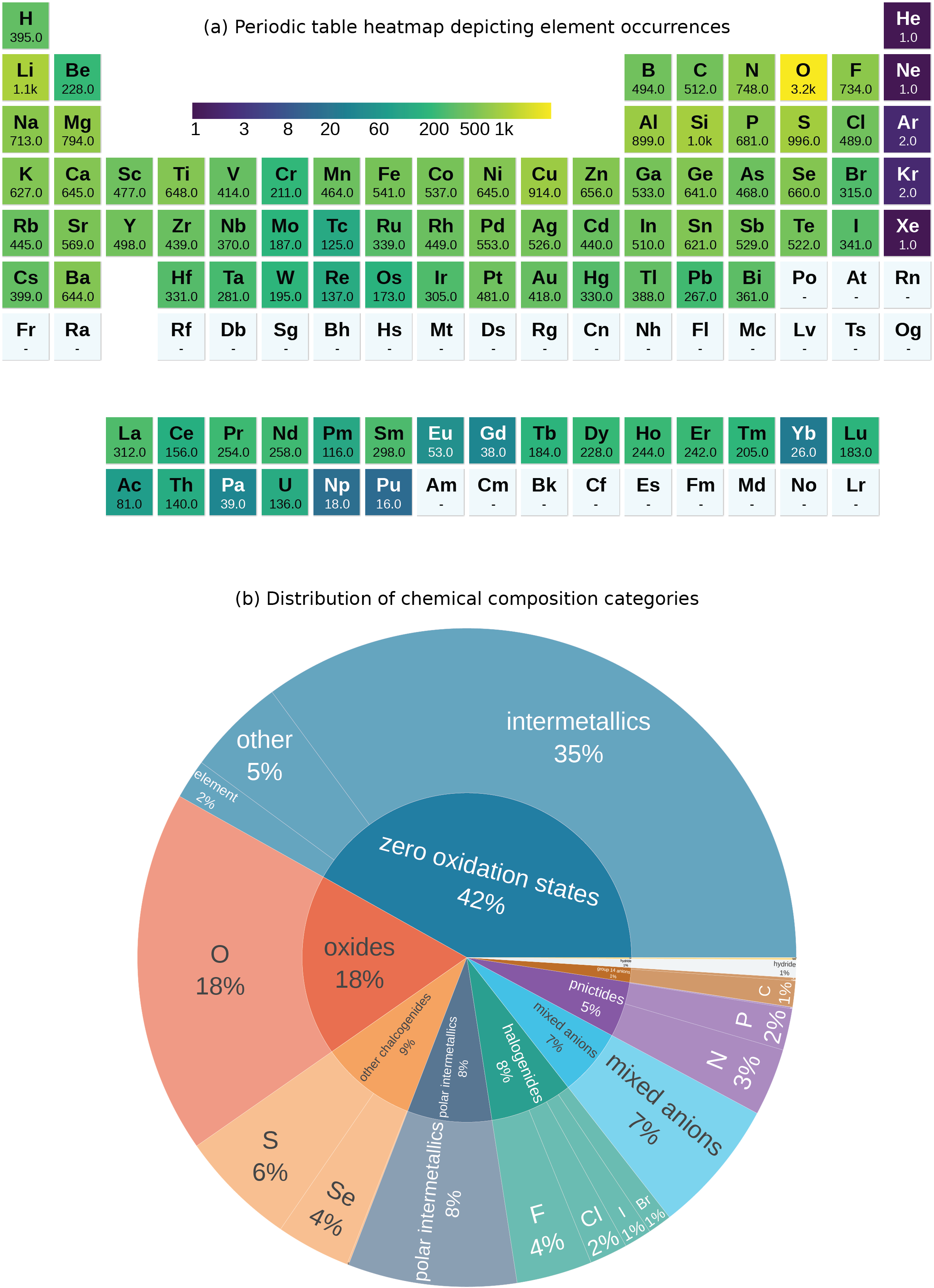}
\caption{Panel (a) presents a logarithmically scaled heatmap of element occurrences across the dataset. Panel (b) illustrates the distribution of chemical composition categories within the dataset.}\label{fig:overview_1}
\end{figure}

\begin{figure}[!ht]
\centering
\includegraphics[width=0.8\linewidth]{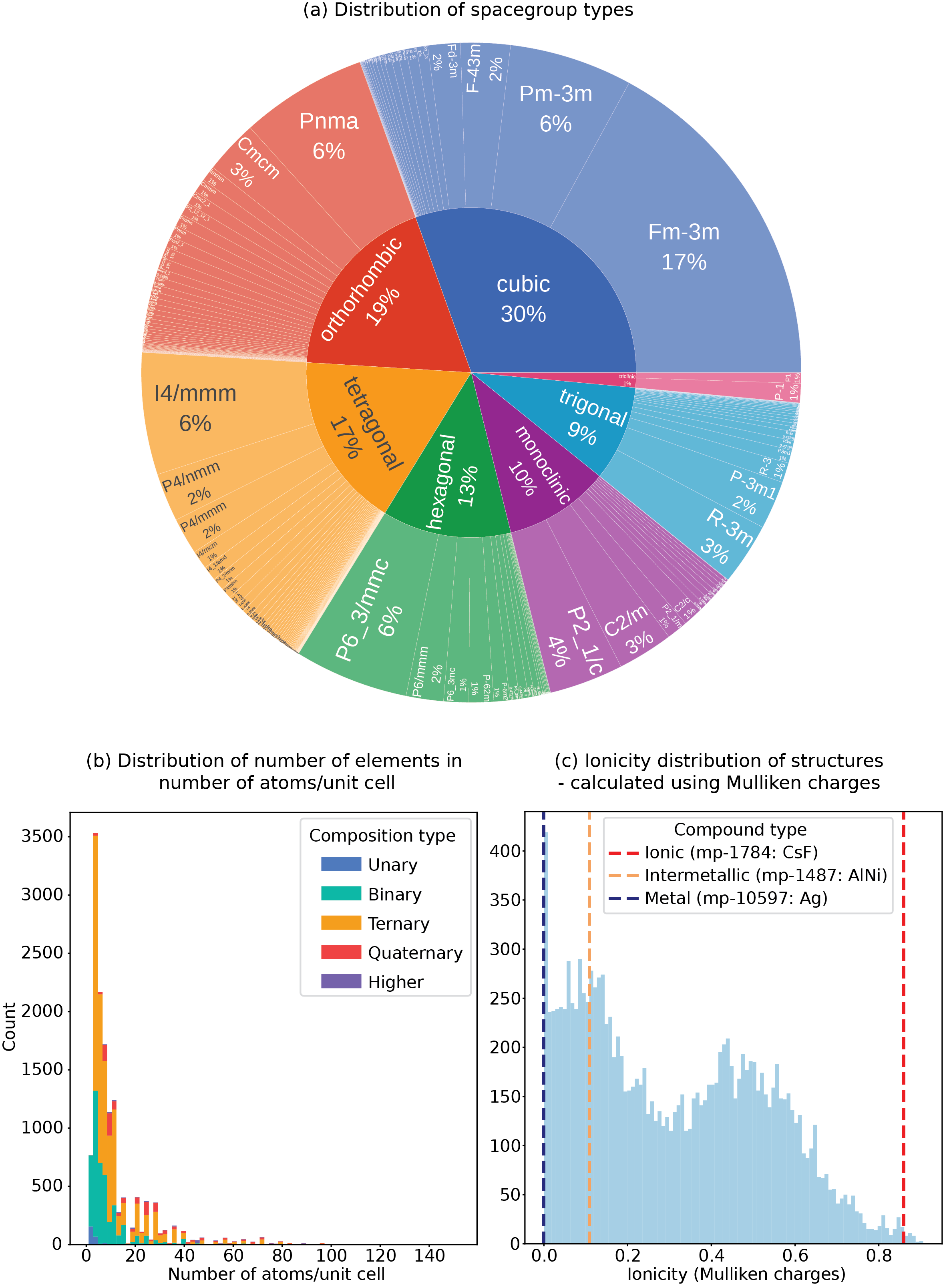}
\caption{Panel (a) shows the distribution of space groups across the dataset. Panel (b) presents the distribution of the number of atoms per unit cell, with colors indicating the number of elements within each bin. Panel (c) illustrates the distribution of calculated ionicity using Mulliken charges from LOBSTER calculations (See equation \ref{eq:S6}) across the dataset for all the structures.}\label{fig:overview_2}
\end{figure}
%\textbf{Supporting Information} \par %Please delete the Suppporting Information statement if it is not applicable. Please supply Supporting Information in another file. Supporting information should not be provided in .tex format
%Supporting Information is available from the Wiley Online Library or from the author.

% References
\medskip

\printbibliography

% Figures/tables and captions
% Permission statements are required for all figures reproduced or adapted from previously published articles/sources. Please also ensure that all necessary permissions to reproduce images have been received
% Please remove these statements for the original figures

% Table of contents entry should be 50 - 60 words long
% Image should be 55 mm broad and 50 mm high or 110 mm broad and 20 mm high

% \newpage
% \begin{figure}[ht]
% \textbf{Table of Contents}\\
% \medskip
%   \includegraphics{toc-image.png}
%   \medskip
%   \caption*{The impact of new bonding descriptors in machine learning models for predicting material properties is assessed. Improvements are validated using significance tests, and new, intuitive descriptors for screening lattice thermal conductivity and projected force constants are introduced. The results of this study not only establish bonding descriptors as a key component but also open new opportunities for further research.}
% \end{figure}

\end{document}